\documentclass[11pt]{article}
\usepackage[english]{babel}
\usepackage{amsmath,amssymb,latexsym,amscd}
\usepackage{theorem}
\usepackage{enumerate}
\usepackage{mjpmath}
\begin{document}
\addcontentsline{toc}{section}{Introduction}
\thispagestyle{empty}
\title{On the deformation quantization of symplectic orbispaces}
\author{Markus J. Pflaum\thanks{Fachbereich Mathematik, 
        Johann Wolfgang Goethe-Universit\"at, 60054 Frankfurt/Main, Germany
        \hspace{10mm}\mbox{ }
        email: {pflaum{\at}math.uni-frankfurt.de}
        }}
\date{\today}
%


%
\maketitle
\begin{abstract}
  In the first part of this article we provide a geometrically 
  oriented approach to the theory of orbispaces which originally 
  had been introduced by Chen. We explain the notion of a vector 
  orbibundle and characterize the good sections of a reduced vector 
  orbibundle as the smooth stratified sections. 
  In the second part of the article we elaborate on the quantizability
  of a symplectic orbispace. By adapting Fedosov's method to the
  orbispace setting we show that every symplectic orbispace has a
  deformation quantization.  As a byproduct we obtain that every symplectic 
  orbifold possesses a star product.
\end{abstract}


%
\tableofcontents
\section*{Introduction}
Deformation quantization has been introduced into mathematical physics
by {\sc Bayen--Flato--Fronsdal--Lichnerowicz--Sternheimer} 
\cite{BayFlaFroLicSte:DTQ} more than 25 years ago. Since then,
various existence and classification results for star products on a symplectic 
or Poisson manifold have appeared 
\cite{WilLec:ESPFDPLAASM,Fed:SGCDQ,Kon:DQPMI}.
A common feature of all these approaches is that the space 
to be quantized is not allowed to have singularities.  
But many symplectic or Poisson spaces with strong relevance for 
mathematical physics
are singular. For instance, the phase spaces appearing in gauge theory or
obtained by symplectic reduction
are in general not smooth and possess singularities.
According to the work of {\sc Sjamaar-Lerman} \cite{SjaLer:SSSR}
such singular symplectically reduced spaces are stratified spaces, where each
stratum carries a canonical symplectic structure.
So the natural question arises, whether an arbitrary symplectic or 
Poisson stratified spaces has a deformation quantization as well.
In this work we consider a particular class of Poisson spaces with 
singularities, namely symplectic orbispaces, 
and show for this class the existence of a star product.
We achieve this by generalizing Fedosov's construction to the orbispace 
setting. 

The category of orbispaces 
has been introduced by {\sc W.~Chen}  \cite{Che:HTO}. 
By definition, orbispaces are topological spaces which locally look like orbit
spaces of compact Lie group actions. Thus orbispaces are a natural 
generalizations of orbifolds and our results imply in particular that 
every symplectic orbifold carries a star product. 

Our article is set up as follows. In the first section we recall the 
notion of a stratification and elaborate on the canonical stratification
of an orbit space by orbit types. Moreover, we introduce profinite
dimensional manifolds and differentiable categories with slices.
Both concepts will be needed later in the
definition of a (possibly infinite dimensional) orbispace.
 
In Section \ref{SecOrbi} we provide an introduction to orbispaces. 
Since the applications we have in mind are of a differential geometric nature
we have adapted the original topological approach 
by {\sc W.~Chen} \cite{Che:HTO} to the differential geometric situation.
Moreover, the approach presented here allows infinite dimensional 
orbispaces. Concerning the subcategory of orbifolds let us mention that  
we do not make any restrictions on the codimension of the fixed point sets 
of the local isotropy groups of the orbifold. This entails in particular
that manifolds with boundary or with corners can be regarded as orbifolds.
In the second part of Section \ref{SecOrbi} we introduce  
the notion of a vector orbibundle and of a reduced resp.~good orbibundle.
The main result is Theorem \ref{ThmGSec}, where we show that a
continuous section of a reduced vector orbibundle is a 
good section in the sense of {\sc Ruan} \cite{Rua:SGTO},
if and only if it is a stratified section which extends to a
(vertical) derivation of the algebra of smooth functions on the orbibundle.
Thm.~\ref{ThmGSec} is  essentially  a consequence of the smooth isotopy
lifting theorem of {\sc Schwarz} \cite{Schwa:LSHOS}.

In the third section we introduce riemannian and symplectic orbispaces.
Moreover, we explain what to understand by a metric resp. symplectic 
connection and show that for every symplectic orbispace there exist
symplectic connections.
The explicit definition what to understand by a deformation quantization
resp.~a star product on a symplectic orbispace is also contained in 
Section \ref{SecSympOrbi}.
In Section \ref{SecFedQOrbi} we construct a star-product on a symplectic
orbispace by localizing Fedosov's method to the orbispace 
charts of an appropriate orbispace atlas. It is a consequence of 
Thm.~\ref{ThmGSec} that this idea works, indeed.
In some more detail, we introduce the formal Weyl algebra 
bundle over a symplectic orbispace and, given a symplectic connection,
construct a flat connection for this bundle.
The fiberwise Weyl--Moyal product on its space of flat sections then 
gives rise to a star product for the symplectic orbispace.     
\vspace{2mm}

{\bf Acknowledgement:} The author gratefully acknowledges financial
support by the European Research Training Network 
{\it Geometric Analysis on Singular Spaces}.
Moreover, the author thanks Jean-Paul Brasselet for the invitation to
the Institut de Math\'ematiques de Luminy, where most of this work has
been done.


%
%
\section{Preliminaries}
\begin{topic} {\bf Stratifications }
In the presentation of the basics of stratification theory we  
follow {\sc Mather} \cite{Mat:SM} 
(see also {\sc Pflaum} \cite[Chap.~1]{Pfl:AGSSS} for further details).

By a {\it decomposition} of a paracompact second countable topological 
Hausdorff space
$X$ one understands a locally finite partition $\cZ$ of $X$ into locally closed
subspaces $S\subset X$ called {\it pieces} such that the following conditions
are satisfied:
\begin{enumerate}[{\rm (DEC1)}]
\item
\label{IteDecSp1}
  Every piece $S\in\cZ$ is a smooth manifold in the induced topology.
\item
\label{IteDecSp2}
  ({\it condition of frontier})
  If $R \cap \clos{S} \neq \emptyset$ for a pair of pieces $R,S\in\cZ$,
  then $R \subset \clos{S}$. In this case one calls $R$ {\it incident} to 
  $S$, or a  {\it boundary stratum} of $S$.
\end{enumerate}
Obviously, the incidence relation is a partial order on the set of pieces.
The set of decompositions of $X$ is partially ordered by the
``coarser''-relation. 
Hereby, a decomposition $\cZ_1$ of $X$ is called {\it coarser} than a 
decomposition $\cZ_2$, if every stratum of $\cZ_2$ is contained in a 
stratum of $\cZ_1$.

By a {\it stratification} of $X$ one now understands a mapping $\cS$ 
which associates to every $x\in X$ the set germ $\cS_x$ of a closed 
subset of $X$ such that the following axiom is satisfied: 
\begin{enumerate}[{\rm (STRA}{\rm )}]
\item
\label{AxStraRa}
  For every $x\in X$ there is a neighborhood $U$ of $x$ and a decomposition
  $\cZ$ of $U$ such that for all $y \in U$ the germ $\cS_y$ coincides
  with the set germ of the stratum of $\cZ$ of which $y$ is an element.
\end{enumerate}
The pair $(X,\cS)$ then is called a {\it stratified space}.
Obviously, a decomposition $\cZ$ induces a stratification of $X$.
The following proposition shows that the converse holds true as well; a
proof of this result can be found in \cite[Prop.~1.2.7]{Pfl:AGSSS}. 
\end{topic} 
\begin{proposition}
\label{Prop1.2}
   Let $\cS$ be a stratification on $X$. Then there exists 
   a coarsest decomposition $\cZ_{\cS}$ of $X$ inducing $\cS$. 
\end{proposition}
We will denote the decomposition $\cZ_\cS$ by $\cS$ as well. 
Its pieces will be called {\it strata}.

\begin{topic} 
\label{ExOrbiType}
{\bf Stratification of orbit spaces }
  Let $G$ be a Lie group acting properly on a smooth manifold $M$. 
  Denote for every compact subgroup $H\subset G$ by $M_H$ 
  the submanifold of all points of $M$ having isotropy group equal to
  $H$ and by $M_{(H)}$ the submanifold of all $x\in M$ having isotropy 
  group conjugate to $H$. If $M_{(H)}\neq \emptyset$, we say that
  the conjugacy class $(H)$ is an {\it orbit type} of $M$.
  The following propositions hold true.
  \begin{enumerate}[(1)] 
  \item 
  \label{StraOrbiIte1}
    If $M/G$ is connected, there exists a compact subgroup 
    $G^\circ \subset G$ such that the subsets
    $M_{(G^\circ)} \subset M$ and $M_{(G^\circ)}/G \subset M/G$ 
    are both open and dense. The set $M_{(G^\circ)}/G$ is connected.
    Moreover, for every $x\in M$ the group $G^\circ$ is conjugate 
    to a subgroup of the isotropy group $G_x$. 
  \item \label{StraOrbiIte2}
    The mapping $\cS$ which associates to every $x$ the set germ
    of $M_{(G_x)}$ is a stratification of $M$. 
    Moreover, the mapping which associates to every orbit $Gx$ 
    the set germ of $M_{(G_x)}/G$ is a stratification 
    of the orbit space $M/G$. 
    The thus defined stratifications are called the stratification of $M$
    resp.~$M/G$ by {\it orbit types}.
  \item
  \label{StraOrbiIte3}
    If $M/G$ is connected, then the  largest normal subgroup of $G$ contained 
    in $G^\circ$ coincides with 
    the kernel of the canonical homomorphism $G \rightarrow \Diff (M)$.
  \item 
  \label{StraOrbiIte4}
    If $G$ is a finite group and $M$ is connected, 
    then $G^\circ$ is a normal subgroup  
    and  $G^\circ \subset G_y$  for every $y\in M$.
    Moreover, $G$ acts effectively on $M$, if and only if 
    $G^\circ$ is trivial.
  \end{enumerate}
\end{topic}
\begin{proof}
   Proposition (\ref{StraOrbiIte1}) is the well-known principal orbit 
   type theorem due to {\sc Montgomery--Samelson--Zippin} 
   \cite{MonSamZip:SPCTG}; see also
   {\sc Bredon} \cite{Bre:ICTG} or \cite[Sec.~4.3]{Pfl:AGSSS} for details.
   A proof of (\ref{StraOrbiIte2}) can be found in  
   {\sc Bierstone} \cite[Chap.~2]{Bie:SOSSEM} or \cite[Sec.~4.3]{Pfl:AGSSS}.
   
   Let us show (\ref{StraOrbiIte3}). To this end consider the 
   canonical  homomorphism $G \rightarrow \Diff (M)$ of $G$ in the 
   diffeomorphism group of $M$. Let $L$ be its kernel. By definition of
   $L$ one has $L\subset g G^\circ g^{-1}$ for all $g\in G$.
   On the other hand, because $(G^\circ)\subset (G_y) $ for all $y\in M$, the
   inclusion $\bigcap_{g\in G} \, g G^\circ g^{-1} \subset L$ holds as
   well, hence $L = \bigcap_{g\in G} \, g G^\circ g^{-1}$.  
   
   Now we come to (\ref{StraOrbiIte4}) and assume that $G$ is finite.
   We will show that the isotropy groups of all $x\in M^\circ$ coincide.
   Clearly, this suffices to prove (\ref{StraOrbiIte4}).
   So let $M^1$ be the stratum of $M$ of codimension $1$. 
   Then $M^\circ \cup M^1$ is a 
   connected open subspace of $M$, as the complement can be decomposed in
   strata of codimension $\geq 2$. According to the slice theorem there 
   exists for  every point
   $x\in M$ an open connected neighborhood $U_x$ which can be mapped by a
   $G_x$-equivariant diffeomorphism onto a $G_x$-invariant open ball
   around the origin of a $G_x$-representation space $E_x$. 
   Now, if $x\in M^\circ$, then every point $z\in U_x$ lies in 
   $M^\circ$ again and has isotropy group equal to $G_x$.
   In case $x\in M^1$, we will consider the representation space $E_x$ to 
   prove that the isotropy groups of all elements of $U_x \cap M^\circ$ 
   coincide. By the slice theorem and the assumptions on $M^1$ the fixed
   point set $E_x^{G_x}$ is a linear subspace of $E_x$ and of codimension $1$.
   Choose a $G_x$-invariant metric $\langle \cdot ,\cdot \rangle$ on $E_x$ and
   let $v$ be a unit vector in the orthogonal complement of $E_x^{G_x}$.
   Then we have $G_x v = \{ v,-v\}$. 
   Let $K \subset G_x$ be the kernel of the map 
   $G_x \ni g \mapsto \langle gv,v\rangle$ and $h$ a group element such that 
   $hv = -v$.
   Then the isotropy group of an element $\lambda v$ with $\lambda >0$ is 
   identical to $K$ and the isotropy group of $-\lambda v$ is given by 
   $hKh^{-1}$. But $hKh^{-1}$
   is equal to $K$, as $K$ is normal. Hence the isotropy groups of all 
   elements of $U_x \cap M^\circ$ coincide.

   Now, as $M^\circ \cup M^1$ is connected, one can connect any two points
   $x,x' \in M^\circ$ by a finite chain of $U_y$ with
   $y\in M^\circ \cup M^1$. In other words this means that there exist
   $y_0,\cdots,y_n \in M^\circ \cup M^1$ such that $y_0=x$, $y_n =x'$ and
   $U_{y_k} \cap U_{y_{k+1}} \neq \emptyset$ for $k\leq n$. By the above
   considerations, the isotropy groups of $x$ and $x'$ then coincide.
   This proves the claim.
\end{proof}
The proof of (\ref{StraOrbiIte4}) entails also the following technical result,
which will be needed later.
\begin{enumerate}[(1)]
\setcounter{enumi}{4}
  \item 
  \label{StraOrbiIte5}
    Let $G$ be finite, $x$ a point of $M^1$, 
    the stratum of codimension $1$, 
    and $U\subset M$  a neighborhood which is $G_x$-equivariantly 
    diffeomorphic to an open ball around the origin of a 
    $G_x$-representation space.
    Then $U\cap M^1$ is connected and $U\cap M^\circ$ has two connected 
    components. 
    Moreover, $G_x$ acts trivially on  $U\cap M^1$, and there exists
    a homomorphism $G_x \rightarrow \Z_2$ with kernel $G^\circ$ 
    such that every element of $G_x \setminus G^\circ$ interchanges the 
    connected components of $U\cap M^\circ$.
\end{enumerate}

\begin{topic}\label{ParPFD}
{\bf Profinite dimensional manifolds}
  A second countable topological Hausdorff space $M$ is called
  a {\it profinite dimensional manifold}, if there exists a projective
  system $(M_i,\mu_{ij})_{i\leq j \in \N}$ of smooth finite dimensional
  manifolds $M_i$ and surjective submersions $\mu_{ij}: M_j \rightarrow M_i$,
  $i\leq j$ such that $M$ coincides with the projective limit, 
  that means 
  \begin{displaymath}
    M = \lim\limits_{\longleftarrow \atop i\in \N} M_i .
  \end{displaymath}
  If $M$ is a profinite dimensional manifold, there exists a unique
  family of continuous surjections 
  $\mu_i : M \rightarrow M_i$ such that $\mu_i = \mu_{ij} \circ \mu_j$
  for all $i\leq j$ and such that $M$ carries the initial topology with
  respect to the $\mu_i$.
  
  By a {\it profinite dimensional vector space} we understand the 
  projective limit
  $ V = \lim\limits_{\longleftarrow \atop i\in \N} V_i $
  of a projective system $(V_i,\varphi_{ij})_{i\leq j \in \N}$
  of  finite dimensional (real) vector spaces $V_i$ and
  surjective linear maps $\varphi_{ij}: V_j \rightarrow V_i$,
  $i\leq j$.  Clearly, every profinite dimensional vector space is
  a profinite dimensional manifold. Examples of profinite dimensional vector
  spaces  are the projective limit
  $\R^\infty = \lim\limits_{\longleftarrow \atop n\in \N} \R^n$ 
  and the completed symmetric tensor algebra 
  $\widehat{\operatorname{Sym}}^\bullet (W):=
  \lim\limits_{\longleftarrow \atop n\in\N} \operatorname{Sym}^\bullet (W)/
  \mathfrak m^n$ of a finite dimensional real vector space $W$. Hereby,
  $\operatorname{Sym}^\bullet (W)$ denotes the (complexified) symmetric 
  tensor algebra of $W$ and $\mathfrak m$ the kernel
  of the canonical homomorphism 
  $\operatorname{Sym}^\bullet (W) \rightarrow \C$.
  Note that $\R^n$ can be naturally embedded as a subspace of 
  $\R^\infty$, since for all $n \leq N$, $\R^n$ is canonically embedded
  in $\R^N$ via the first $n$ coordinates.

  The {\it sheaf of  smooth functions} on a profinite dimensional manifold 
  $M = \lim\limits_{\longleftarrow \atop i\in \N} M_i $ 
  is defined as the sheaf  $\cC^\infty_M$ with sectional spaces
  \begin{displaymath}
    \cC^\infty_M (U) =\{ g \in \cC (U) \mid \text{there exist $i\in \N$ 
    and  $g_i \in \cC^\infty (\mu_i (U))$ such that 
    $g_i \circ {\pi_i}_{|U} = g$} \},
  \end{displaymath}
  where $U$ runs through the open subsets of $M$.
  Given a second profinite dimensional manifold 
  $N = \lim\limits_{\longleftarrow \atop i\in \N} N_i $, a continuous map 
  $f: M \rightarrow N$ is called {\it smooth},
  if $f_* \cC^\infty_M \subset \cC^\infty_N$. 
  Using Whitney's embedding theorem it is straightforward to check that
  for every smooth map $f: N \rightarrow M$ there exists,
  possibly only after passing to projective subsystems of $(M_i,\mu_{ij})$ and
  $(N_i,\nu_{ij})$, a family of smooth maps 
  $f_i:N_i\rightarrow M_i$ such $f_i \circ \nu_i = \mu_i \circ f$ for all $i$.
  In case the $f_i$ can be chosen to be immersions 
  (resp.~submersions, embeddings or diffeomeorphisms), one says that
  $f$ is an immersion (resp.~submersion, embedding or diffeomeorphism).
  Using Whitney's embedding theorem again one proves that every 
  profinite dimensional $M$ can be embedded in 
  $\R^\infty$.  
  
  Obviously, the profinite dimensional manifolds and the smooth maps between
  them form a category which we will denote by 
  $\mathfrak{Man}_{\text{\sf \tiny pf}}$.
  Similarly, the profinite dimensional vector spaces with smooth
  linear maps as morphisms form a category.

  If a compact Lie group $G$ acts on a profinite dimensional manifold
  $M = \lim\limits_{\longleftarrow \atop i\in \N} M_i $, one can construct
  a $G$-invariant riemannian  metric on $M$. Given a point $x\in M$, 
  such a riemannian metric gives rise to a $G$-invariant tubular neighborhood
  of the orbit through $x$. From this one concludes by a standard argument
  that the slice theorem holds as well for compact Lie group actions on 
  profinite dimensional manifolds.
\end{topic}

\begin{topic} 
\label{TopTCFS}
{\bf Differentiable categories with slices}
Consider a subcategory of the category 
of profinite dimensional manifolds and smooth maps.
In this article we will denote such a subcategory by $\mathfrak{T}$ and
always assume that it satisfies the following axioms.
\begin{enumerate}[({DCAT}1)]
\item
\label{DCAT1} 
 For every morphism $f:N \rightarrow M$ in $\mathfrak{T}$ which is a
 smooth open embedding of profinite dimensional manifolds,
 the image $f (N)$ is an open subobject of $M$. 
 $U\subset M$ being an open subobject
 hereby means that $U$ is open and that the canonical injection 
 $U\hookrightarrow M$ is a morphism in $\mathfrak{T}$.
\item
\label{DCAT2}
 For every object $M$ the set of open subobjects is a topology on $M$.
\end{enumerate}
With a view towards symmetries  we assume additionally that $\mathfrak T$
satisfies the axiom (SLC) below; a category for which 
(DCAT\ref{DCAT1}), (DCAT\ref{DCAT2}) and (SLC) are true will be called a
{\it differentiable category with slices}.
\begin{enumerate}[(SLC)]
\item 
\label{Slc}
  Let $(M,G)$ be an object of $\mathfrak{T}^{\text{\sf \tiny sym}}$
  and $x\in M$ a point.
  Then there exists a $\mathfrak{T}$-slice for $M$ at $x$ that means 
  an embedding $(\xi,\lambda):(S,K)\rightarrow (M,G)$ with $\lambda$
  injective and a point $s\in S$ such that $\xi (s) =x$  
  and such that $(\xi,\lambda)$ is universal in the following sense.
  Assume to be given an embedding $(\varphi,\iota):(N,H)\rightarrow (M,G)$ 
  and a point $y\in N$ where $\iota$ is injective and 
  $x=\varphi (y)$. Then there exists, after passing to appropriate open
  subobjects, an equivariant automorphism 
  $(\Phi,\id):(M,G)\rightarrow (M,G)$ with $\clos{\Phi} =\id_{M/G}$ 
  and an embedding $( \psi,\kappa): (S,K)\rightarrow (N,H)$ 
  such that $\psi (s) =y$ and such that the following diagram commutes: 
  \begin{equation}
    \begin{CD}
      (S,K)@>_{(\xi,\lambda)}>> (M,G) \\
      @V{(\psi,\kappa)}VV @VV{(\Phi,\id)} V\\
      (N,H) @>_{(\varphi,\iota)}>> (M,G).
    \end{CD}
  \end{equation}
\end{enumerate}

As typical examples for a differentiable category with slices we have
the following in mind; using the slice theorem the reader will easily check 
that these categories satisfies the above axioms and in particular 
(SLC): 
\begin{enumerate}[(1)]
  \item
        the category $\mathfrak{Man}$
        of finite dimensional smooth manifolds and smooth maps, 
  \item
        the category $\mathfrak{Man}_{\text{\sf \tiny pf}}$
        of profinite dimensional manifolds and smooth maps,
  \item
        the category $\mathfrak{VBdl}$ 
        of smooth vector bundles over finite dimensional manifolds; 
        hereby the fiber vector space is allowed 
        to be a  profinite dimensional vector space and 
        the morphisms are given
        by smooth vector bundle maps over smooth maps 
        between the bases.         
\end{enumerate}

The category $\mathfrak{T}^{\text{\sf \tiny sym}}$ of
$\mathfrak{T}$-{\it objects with {\rm (}compact{\rm )} symmetries}
consists of the following 
object and morphism classes. Objects are given by pairs $(M,G)$, where
$M \in \Obj (\mathfrak{T})$ and $G$ is a compact Lie group which acts 
smoothly on $M$ by elements of 
the automorphism group $\Aut_{\mathfrak{T}} (M)$.
Morphisms are  given by equivariant maps
$(\varphi,\iota ): (N,H) \rightarrow (M,G)$. This means that
$\iota :H \rightarrow G$ is a continuous group homomorphism and 
$\varphi :M\rightarrow N$ a morphism of $\mathfrak{T}$ such that
$\varphi(hy) =\iota(h)\, \varphi (y)$ for all $y\in N$ and $h\in H$.
Two equivariant maps
$(\varphi,\iota), (\varphi',\iota'): (N,H) \rightarrow (M,G)$
are said to be {\it equivalent}, if there exists an element $g\in G$ 
such that $(\varphi',\iota') = (g,\operatorname{Ad}_g)(\varphi,\iota)$.

The following properties of a morphism 
$(\varphi,\iota):(N,H) \rightarrow (M,G)$ in 
$\mathfrak{T}^{\text{\sf \tiny sym}}$ are easy to prove:
\begin{enumerate}[(SYM1)]
\setcounter{enumi}{-1}
\item
\label{SYM0} 
  $\varphi$ induces a continuous map 
  $\overline{\varphi}: N/H \rightarrow M/G$ between orbit spaces.
\item
\label{SYM1}
  If $\varphi$ is surjective and $G$ acts effectively on $M$, then $\iota$ is
  uniquely determined by $\varphi$.
\item
\label{SYM2}
  If $\varphi$ is injective and $H$ acts effectively on $N$, then $\iota$ is a
  monomorphism. 
\end{enumerate}
Let us introduce some useful notation.
An object $(M,G)$ of $\mathfrak{T}$ is called {\it reduced}, if
$G$ acts effectively on $M$. Note that for arbitrary $(M,G)$ there
exists a natural equivariant morphism from $(M,G)$
onto the reduced object $(M,G_{M,\text{eff}})$, where 
$G_{M,\text{eff}}$ is the quotient group of $G$ by the kernel of the 
homomorphism $G\rightarrow \Aut_{\mathfrak{T}} (M)$. 

A morphism $(\varphi,\iota): (N,H) \rightarrow (M,G)$ between objects of 
$\mathfrak{T}^{\text{\sf \tiny sym}}$ is called an {\it embedding}, 
if $\varphi$ is a smooth embedding and $\overline{\varphi}$ a 
homeomorphism onto an open subset of the orbit space $M/G$. 
If additionally $\varphi$ is an open map, 
we say that $(\varphi,\iota)$ is an {\it open embedding}. 
Note that for 
$(\varphi,\iota)$ an embedding, $\iota$ need not be a monomorphism.
Moreover, (SLC) implies that for every object $(M,G)$
of $\mathfrak{T}^{\text{\sf \tiny sym}}$ and every point $x\in M$ there
exists an embedding $(\varphi,\id_H) : (S,H) \rightarrow (M,G)$.

The  following further properties hold for
finite symmetries in  a differentiable slice category $\mathfrak{T}$.
\begin{enumerate}[(SYM1)]
\setcounter{enumi}{2}
\item  
\label{SYM3}
 Assume that $N$ is connected and that $G,H$ are finite. 
 Let $(\varphi,\iota)$ and $(\varphi',\iota')$ be
 two open embeddings from $(N,H)$ to $(M,G)$ with the actions of
 $G$ and $H$ effective. 
 Then  $(\varphi,\iota)$ and $(\varphi',\iota')$ are equivalent, if and
 only if $\overline{\varphi}=\overline{\varphi'}$.
\item
\label{SYM4}
 Assume that $N$ is connected and that $G,H$ are finite.
 Let  $(\varphi,\iota) :(N,H) \rightarrow (M,G)$ be
 an open embedding and assume that $G$ acts effectively on $M$. Then, if
 $g \varphi (N)\cap \varphi (N) \neq \emptyset$ for $g\in G$, the relation
 $g \varphi (N)= \varphi (N)$ holds true and $g$ lies in the image of $\iota$. 
\end{enumerate}
\end{topic}
\begin{remark}
  (SYM\ref{SYM3}) and (SYM\ref{SYM4}) correspond to Lemma 1 and 
  Lemma 2 in {\sc Satake} \cite{Sat:GBTVM}, but note that
  in \cite{Sat:GBTVM} the additional assumption has been made that 
  $(M,G)$ and $(N,H)$ do not contain strata of codimension $1$. 
  In the following we repeat Satake's short proof of (SYM\ref{SYM4}), 
  which also works in the general case of strata
  of arbitrary codimension, and provide a new argument showing
  that (SYM\ref{SYM3}) is true without any assumptions on the
  codimension of the strata. 
\end{remark}
\begin{proof} 
   Let us first prove the claim for the case where $\mathfrak{T}$ is the 
   category of (finite dimensional) smooth manifolds and smooth maps.
   Denote by $M^\circ$ the open stratum of a $G$-manifold $M$ and 
   by $M^1$ the stratum of codimension $1$ with respect to the stratification 
   by orbit types. Likewise define $N^\circ$ and $N^1$ for an
   $H$-manifold $N$.
   Now, we will show first property (SYM\ref{SYM4}) and afterwards 
   (SYM\ref{SYM3}). 
 
   So assume that $N$ is connected, $(\varphi,\iota)$ is an open
   embedding and that
   $g \varphi (N) \cap \varphi (N)\neq \emptyset$. Then there
   exist $y,y'\in N^\circ$ such that $\varphi (y) \in M^\circ$ and
   $\varphi (y') = g \varphi (y) $. 
   As $\overline \varphi$ is injective, $y$ and $y'$ have to ly in the 
   same $H$-orbit, hence  $y' = hy$ for some $h\in H$. 
   We then have $ \varphi (h z)= g' \varphi (z)$ for all $z\in N$ and
   $g'=\iota (h)$. As $\varphi (y) \in M^\circ$ and $G$ acts effectively,
   we have $g=g'=\iota (h)$ and consequently
   $g\varphi (N) =  \varphi (hN) = \varphi (N)$. 
   This shows (SYM\ref{SYM4}).  

   Next we consider (SYM\ref{SYM3}). 
   Assume that $\varphi' (y_\circ ) = \varphi (y_\circ)$ 
   for some $y_\circ \in N^\circ$.  We will then show that $\varphi'=\varphi$
   and $\iota'= \iota$. Clearly, this will prove (SYM\ref{SYM3}).
   Using (SYM\ref{SYM4}) it is straightforward to check that 
   $\varphi (N^\circ) \subset M^\circ$ and 
   $\varphi' (N^\circ) \subset M^\circ$.
   Let us prove that $\varphi (N^1) \subset M^1$. 
   To this end choose for every point $y\in Y$ 
   an $H_y$-invariant neighborhood $V_y$ such that 
   $hV_y \cap V_y= \emptyset$ for $h \in H\setminus H_y$ and such that $V_y$ is
   equivariantly diffeomorphic to an $H_y$-invariant open ball around the 
   origin in a linear $H_y$-representation space. 
   In case $y\in N^1$, we know by  \ref{ExOrbiType} (\ref{StraOrbiIte5})
   that $V_y\cap  N^1$ is connected, that $V_y\cap  N^\circ$ has two connected 
   components and that $H_y \cong \Z_2$. Hence, by (SYM\ref{SYM2})
   $\Z_2\cong \iota (H_y) \subset G_{\varphi (y)}$.
   The subgroup $\iota (H_y)$ acts trivially on the 
   manifold $U_{\varphi (y)}^1 := \varphi (V_y\cap N^1)$, 
   and the nonneutral element interchanges the connected
   components  of $\varphi (V_y\cap N^\circ)$. As a consequence of
   \ref{ExOrbiType} (\ref{StraOrbiIte4}), 
   $G_{\varphi (y)}$ acts effectively on a 
   neighborhood of $\varphi (y)$ contained in
   $U_{\varphi (y)} := \varphi (V_y)$.  
   So, if $\iota (H_y) \neq G_{\varphi (y)}$, one can
   find by (SYM\ref{SYM4}) an element  
   $k\in G_{\varphi (y)} \setminus \iota (H_y)$ and a point 
   $x\in U_{\varphi (y)} $ with $kx \in  U_{\varphi (y)}$
   and $kx \notin \iota (H_y) x $. But this contradicts the fact that
   $\overline\varphi$ is injective. Hence $G_{\varphi (y)} \cong \Z_2$
   and consequently $\varphi (N^1)\subset M^1$. The same argument also 
   proves $\psi (N^1)\subset M^1$.
   We continue with the proof of the equality  
   $\varphi' = \varphi$.  Let $A$ be the set 
   $\{ y\in N \mid \varphi' (y) = \varphi (y)\}$.
   Obviously, $A$ is  closed in $N$ and nonempty, since $y_\circ\in A$.
   Let us show that $A\cap N^\circ$ is also open.
   Let $y\in A\cap N^\circ$ and assume that there exists 
   a sequence $(y_n) \subset N^\circ \setminus A$ converging
   to $y$. After transition to an appropriate subsequence there exists 
   $g' \neq e$ such that $\varphi' (y_n) = g' \varphi (y_n)$ for all $n$.
   By continuity $\varphi' (y) = g' \varphi (y)$ follows, hence 
   $ \varphi (y) = g' \varphi (y)$.
   But this contradicts $G_{\varphi (y)} =\{ e\}$, so
   $A\cap N^\circ$ must be open indeed. 
   Now let $y\in N^1$ and assume that 
   $A \cap V_y \cap N^\circ \neq \emptyset$.
   \ref{ExOrbiType} (\ref{StraOrbiIte5}) 
   entails that $V_y$ can be decomposed in three connected 
   subsets $V_y^{\text{\tiny N}}$, $V_y^{\text{\tiny S}}$ and $V_y^1$,
   where the first two are the connected components of 
   $V_y \cap N^\circ$ and $V_y^1$ is equal to $V_y \cap N^1$.
   By assumption on $y$ there exists $z_0 \in V_y \cap N^\circ$, 
   let us say $z_0 \in V_y^{\text{\tiny N}}$, such that  
   $\varphi' (z_0) =  \varphi (z_0)$. 
   By the results proven so far we know that $\varphi' (z) = \varphi(z)$ 
   for all $z \in V_y^{\text{\tiny N}} \cup V_y^1$.
   We now want to show that this holds for $z \in V_y^{\text{\tiny S}}$
   as well. As it has been shown above, both $H_y$ and 
   $G_{\varphi (y)}$ are isomorphic to $\Z_2$.
   Let $h$ be the nonneutral element of $H_y$.
   Then  both $\iota (h)$ and $\iota' (h)$ coincide with the nonneutral 
   element of $G_{\varphi (y)}$; this implies in particular 
   that $\iota' (h) =  \iota (h)$.
   As $hz \in  V_y^{\text{\tiny N}}$ 
   for $z \in V_y^{\text{\tiny S}}$, we obtain 
   \begin{displaymath}
     \varphi' (z) = \iota' (h) \varphi' (h^{-1}z ) = 
     \iota (h) \varphi (h^{-1}z)  =  \varphi (z),
   \end{displaymath}
   hence $\varphi' (z) = \varphi(z)$ for all $z\in V_y$. 
   Since every element of $N^\circ \cup N^1$ can be connected with $y_\circ$ 
   by a finite chain of $V_y$ with either $y\in N^\circ$ or $y\in N^1$, 
   this shows that $N^\circ \cup N^1$ is contained in $A$. 
   As $A$ is closed and $N^\circ$  is dense in $N$, 
   we thus obtain $A = N$. This proves the relation $\varphi' = \varphi$
   under the assumption of finite $G$ and $H$.
   To show that $\iota' =  \iota $ consider the open set 
   $V=H\, V_{y_\circ}\subset N^\circ$ and the image 
   $U= \varphi' (V) =  \varphi (V)$. Obviously, 
   $\im\iota'\subset G_U := \{\tilde g\in G\mid \tilde gU \subset U\}$. 
   Since  $\varphi' (hy) =  \varphi (hy) = \iota (h)  \varphi' (y)$
   for $y\in V$ and as $G_U$ acts effectively on $U$, the relation
   $\iota' =  \iota $ follows.
   This finishes the proof of axiom (SYM\ref{SYM3}).  

   For the case of profinite dimensional manifolds with finite symmetries
   $(M = \lim_{\longleftarrow \atop i\in \N} M_i,G)$ 
   and $(N=\lim_{\longleftarrow \atop i\in \N} N_i,H)$ 
   one concludes the claim from the fact that 
   axioms (SYM\ref{SYM3}) and  (SYM\ref{SYM4}) hold true for the 
   components $(M_i,G)$  and $(N_i,H)$. The details of the 
   corresponding straightforward argument are left to the reader.
   Finally, an arbitrary differentiable slice category $\mathfrak{T}$ 
   satisfies
   (SYM\ref{SYM3}) and  (SYM\ref{SYM4}) since these axioms hold true
   for $\mathfrak{Man}_{\text{\sf \tiny pf}}$.
\end{proof}


%
%
\section{Orbispaces}
\label{SecOrbi}
\begin{topic} {\bf Orbispace charts }
Let $X$ be a topological Hausdorff space and 
$\mathfrak{T}$ a differentiable category.
By a $\mathfrak{T}$-{\it orbispace chart} for $X$ we understand 
a triple $(\tilde U,G,\varrho)$ such that $(\tilde U,G)$ is an object of
$\mathfrak{T}^{\text{\tiny \sf sym}}$ and 
$\varrho: \tilde U \rightarrow U\subset X$
a continuous $G$-invariant map inducing a homeomorphism
$\overline\varrho: \tilde U / G \rightarrow U$ onto an open subset
of $X$. The set $U$ will be called the {\it image} of the orbispace
chart, $\tilde U$ its {\it domain}.
In case the symmetry group $G$ is finite, $(\tilde U, G,\varrho)$
is called a $\mathfrak{T}$-{\it orbifold chart} for $X$. 
A {\it morphism} 
between two  $\mathfrak{T}$-orbispace charts 
$(\tilde V,H,\upsilon)$ and $(\tilde U,G,\varrho)$ 
is a morphism 
$(\varphi,\iota):(\tilde V,H)\rightarrow (\tilde U,G)$ in 
$\mathfrak{T}^{\text{\tiny \sf sym}}$ 
such that $\varrho\circ \varphi = \upsilon$.  
Note that for every $\mathfrak{T}$-orbispace chart $(\tilde U,G,\varrho)$ the 
triple $(\tilde U,G_{\tilde U,\text{eff}},\varrho)$ is a 
$\mathfrak{T}$-orbispace chart as well. 
If $(\tilde U,G,\varrho) = (\tilde U,G_{\tilde U,\text{eff}},\varrho)$,  
we say that $(\tilde U,G,\varrho)$ is a {\it reduced} orbispace chart.
The category of all $\mathfrak{T}$-orbispace charts for $X$ will be 
denoted by $\mathfrak{T}^{\text{\sf \tiny sym}}_X$.

Two $\mathfrak{T}$-orbispace charts 
$(\tilde U_1,G_1,\varrho_1)$ and $(\tilde U_2,G_2,\varrho_2)$
are called {\it germ equivalent} 
at a point $x\in U_1\cap U_2$, if there exist two embeddings
$(\varphi_i,\iota_i):(\tilde V,H,\upsilon)\rightarrow 
(\tilde U_i,G_{i},\varrho_i)$, $i=1,2$, and a distinguished point 
$\tilde x\in \tilde V$ such that $\varphi_i (\tilde V)$ is a 
subobject of $\tilde U_i$ and such that $\upsilon (\tilde x) = x$.
In other words, germ equivalency of orbispace charts means essentially
that the slices of 
$\tilde U_1$ at some point $\tilde x_1 \in \varrho_1^{-1} (x)$ and
of $\tilde U_2$ at some point $\tilde x_2 \in \varrho_2^{-1} (x)$  
coincide (up to isomorphy).
Using axiom (SLC) it is straightforward to check that the germ equivalence 
of orbispace charts at a point $x\in X$ is an equivalence relation indeed.
By a $\mathfrak{T}$-{\it orbispace atlas} for $X$ we now understand a 
covering of $X$ by $\mathfrak{T}$-orbispace charts such that any two of the 
orbispace charts are germ equivalent at every point of the
intersection of their images. If every element of an orbispace atlas is a
$\mathfrak{T}$-orbifold chart, we call the atlas a
$\mathfrak{T}$-orbifold atlas. Obviously, the set of 
$\mathfrak{T}$-orbifold atlases for $X$ is partially 
ordered by inclusion, and for every $\mathfrak{T}$-orbifold atlas $\cA$
there exists a unique maximal $\mathfrak{T}$-orbifold
atlas $\cA_{\text{max}}$ containing $\cA$. Clearly, the same holds for
orbispace atlases.
We arrive at the definition of a $\mathfrak{T}$-{\it orbifold};
this is just a second countable paracompact topological Hausdorff space $X$ 
together with a maximal $\mathfrak{T}$-orbifold atlas, usually denoted by 
$\cA_X$. If $\mathfrak{T}$ is the category of finite dimensional 
manifolds (resp.~profinite dimensional manifolds), a
$\mathfrak{T}$-orbifold is briefly called an orbifold 
(resp.~profinite dimensional orbifold).

Particularly convenient for the study of orbifolds are the so-called 
{\it linear  orbifold charts}. These are orbifold charts 
$(\tilde W, G, \varrho )$, where $\tilde W$ is an open convex 
neighborhood of the origin of some finite dimensional 
$G$-representation space. In this situation we sometimes say that
$x = \varrho (0) \in W$ is the {\it center} of $(\tilde W, G, \varrho)$  
or that $(\tilde W,G,\varrho)$ is {\it centralized} at $x$.
By the slice theorem it is clear that every orbifold germ at 
$x$ can be represented by a linear orbifold chart centralized at this point.

\end{topic}
\begin{topic} {\bf Orbispace functors }
 Let $\cU$ be an open covering of $X$ and $\clos{\cU}$ 
 the category whose objects are given by connected components of finite 
 intersections $U_1\cap \ldots \cap U_k$ of elements $U_1,\ldots,U_k \in \cU$ 
 and whose morphisms are the canonical inclusions.  
 By a $\mathfrak{T}$-{\it  orbispace functor} we understand a
 functor $\mathsf X$ defined on $\clos{\cU}$ and with values in the category
 of orbispace charts of $X$ such that the following conditions hold
 true:
 \begin{enumerate}[(OSF1)]
 \item 
 \label{OSF1}
   For every object $U$ of $\clos{\cU}$ the orbispace chart
   $\mathsf X (U)$ has image $U$.
 \item 
 \label{OSF2}
   The domain $\tilde U$ of every orbispace chart
   $\mathsf X (U)$, $U \in \clos{\cU}$ is connected.
 \item
 \label{OSF3} 
   For all objects  $U,V$ of $\clos{\cU}$ with $V\subset U$ 
   the morphism 
   $\mathsf X_{VU} := \mathsf X (V \rightarrow U) $ 
   is an open embedding.
 \end{enumerate}
A $\mathfrak{T}$-{\it orbispace} now is a second countable and locally 
connected paracompact topological Hausdorff space $X$ together with a  
$\mathfrak{T}$-orbispace functor 
$\mathsf X:\clos{\cU}\rightarrow\mathfrak{T}^{\text{\sf \tiny sym}}_X$.
Clearly, this functor uniquely determines a maximal atlas $\cA_X$ of 
orbispace charts such that 
$\mathsf X $ has image in $\cA_X$. From now on only the elements of $\cA_X$ 
will be called orbispace charts for the $\mathfrak{T}$-orbispace $X$.

If $\mathfrak{T}$ is the category of finite dimensional 
manifolds (resp.~profinite dimensional manifolds), we use the same
language like for orbifolds and briefly say orbispace 
(resp.~profinite dimensional orbispace) instead of
$\mathfrak{T}$-orbispace.
\end{topic}

Using the paracompactness of an orbifold $X$, the following result 
can be easily derived from (SYM\ref{SYM3}) and (SYM\ref{SYM4}). 
We leave the details to the reader. 
\begin{proposition}
  For every $\mathfrak{T}$-orbifold $X$ there exists a
  $\mathfrak{T}$-orbispace functor 
  $\mathsf X:\clos{\cU} \rightarrow \cA_X
  \subset \mathfrak{T}^{\text{\sf \tiny sym}}_X$.
\end{proposition}

\begin{topic} {\bf Stratification of orbispaces } 
  Every orbispace $X$ has a canonical stratification. 
  To construct this consider a point $x$ and choose an 
  orbispace chart 
  $(\tilde U , G, \varrho)$ around $x$. 
  Denote by $\cS_x$ the set germ at $x$ 
  of the stratification of $U \cong \tilde U/G$ by orbit types (recall 
  Example \ref{ExOrbiType}). As a consequence of the slice theorem,
  $\cS_x$ does not depend on the particular choice of 
  $(\tilde U , G, \varrho)$.
  Since the set germ $\cS_x$ is locally induced by a decomposition, 
  we thus obtain a stratification $\cS$, called the 
  {\it canonical stratification} of the orbispace. 
  Proposition \ref{Prop1.2} guarantees the existence of a canonical 
  decomposition of $X$ into smooth manifolds, called the {\it strata} of 
  the orbispace. Moreover, if $X$ is connected, there exists an open and 
  dense stratum which coincides with the regular part of $X$
  and which will be denoted by $X^\circ$.
  The {\it dimension} of $X$ is defined as the dimension of $X^\circ$.
\end{topic}

\begin{example}
  Every manifold with boundary $M$ carries in a natural way the structure of 
  a finite dimensional orbifold. To see this choose a smooth collar
  $c: \partial M \times [0,1) \rightarrow M$, denote by
  $\tilde U_\circ $ the interior $M\setminus \partial M$ 
  and put $\tilde U_1 = \partial M \times (-1,1)$. Then 
  $\Z_2$ acts on $\tilde U_1$ by $(p,t,\pm 1) \mapsto (p,\pm t)$,
  and the map $\varrho_1: \tilde U_1 \rightarrow \im c$, 
  $(p,t) \mapsto c(p,t^2)$
  induces a homeomorphism $\tilde U_1/ \Z_2 \rightarrow \im c$.
  It is now immediate to check that $(\tilde U_\circ, \{ e\}, \id)$ 
  and $(\tilde U_1,\Z_2,\varrho_1)$ comprise an orbifold atlas for $M$. 
  Similarly, though technically somewhat more involving, one proves that
  every manifold with corners is naturally a finite dimensional
  orbifold.

  Note that in the approach to orbifolds going back to 
  {\sc Satake} \cite{Sat:GNM}, manifolds with boundary or corners
  are not regarded as orbifolds (or better  V-manifolds in the language
  of \cite{Sat:GNM}), since every orbifold chart around a boundary
  point possesses a stratum of codimension $1$.
\end{example}

\begin{topic} 
  Given an open  covering $\cU$ of some locally connected topological
  space $Y$, any faithful functor 
  $\mathsf Y:\clos{\cU}\rightarrow \mathfrak{T}^{\text{\sf \tiny sym}}$
  which satisfies axioms (OSF\ref{OSF2}) and (OSF\ref{OSF3})
  above will be called a $\mathfrak{T}$-{\it orbispace functor}, too.
  Hereby, faithful means that the image 
  $\mathsf Y_{vu} ( \mathsf Y (v))$ is properly contained in $\mathsf Y (u)$
  for all $v,u \in \clos{\cU}$ with $v \subsetneq u$.
  The following proposition shows that this new notation 
  is justified indeed.
\end{topic}   
 
\begin{proposition}
\label{Prop2.8}
  Let $\mathsf Y:\clos{\cU}\rightarrow\mathfrak{T}^{\text{\sf\tiny sym}}$
  be a faithful functor satisfying axioms (OSF\ref{OSF2})
  and (OSF\ref{OSF3}). 
  Then there exists a $\mathfrak{T}$-orbispace $X$, 
  an order preserving injective map $\mathsf U$ from $\clos{\cU}$ to the
  topology of $X$ and a $\mathfrak{T}$-orbispace functor
  $\mathsf X: \mathsf U (\clos{\cU}) 
  \rightarrow\mathfrak{T}^{\text{\sf\tiny sym}}_X$,
  $ u \mapsto (\tilde U_u,G_u,\varrho_u)$ 
  such that 
  $\mathsf Y = \mathsf F \circ \mathsf X \circ \mathsf U$,
  where $\mathsf F :\mathfrak{T}^{\text{\sf\tiny sym}}_X \rightarrow
  \mathfrak{T}^{\text{\sf\tiny sym}}$ is the forgetful functor 
  $(\tilde U,G,\varrho)\mapsto (\tilde U,G)$. 
  Moreover, these objects are unique up to isomorphy in the sense that 
  if $X'$, $\mathsf U'$ and $\mathsf X'$ also have this property,
  then there exists a homeomorphism $f: X\rightarrow X'$ 
  such that $\mathsf U' = f\circ \mathsf U$ and
  $\varrho'_{u}= f\circ \varrho_{u}$ for all $u\in \clos{\cU}$.
\end{proposition}
\begin{proof}
  To construct $X$,  $\mathsf X$  and $\mathsf U$ let us first
  denote every object $\mathsf Y (u)$, $u\in \clos{\cU}$ by 
  $(\tilde U_u,G_u)$ and every morhism $\mathsf Y_{vu}$ for $v\subset u$
  by $(\varphi_{vu},\iota_{uv})$.
  Then put
  \begin{displaymath}
    X := \bigsqcup\limits_{u\in \clos{\cU}}\,
    \tilde U_u /G_u \, \Big/ \, \sim ,
  \end{displaymath}
    where two points $x\in \tilde U_u / G_u$ and 
    $x'\in \tilde U_{u'} / G_{u'}$ are in relation $\sim$, if
    there exists $v\in\clos{\cU}$ and a point $y\in \tilde U_v /G_v$ 
    such that  $x=\overline\varphi_{vu}(y)$ and
    $x'=\overline\varphi_{vu} (y)$. The set $X$ carries a natural
    topology given by the quotient topology from the (disjoint) 
    topological sum of the orbit spaces $\tilde U_u /G_u$.
    Now let $\varrho_u$ be the natural map from $\tilde U_u$ to $X$,
    denote by $U_u$ the image of $\varrho_u$, and let $\mathsf X$ be the
    functor $u \mapsto (\tilde U_u,G_u,\varrho_u)$,
    $(v\rightarrow u) \mapsto (\varphi_{vu},\iota_{uv})$.
    Finally define $\mathsf U$ by $\mathsf U(u) =U_u$. Then the objects
    $X$, $\mathsf X$ and $\mathsf U$ satisfy the claim of the proposition. 
    The proof of
    uniqueness up to isomorphy is given by standard arguments, so 
    we will  leave it to the reader.
\end{proof}
   In the situation of the proposition we say that the $\mathfrak{T}$-orbispace
   $X$ is {\it induced} by $\mathsf Y$. For convenience we will also 
   notationally identify the functors $\mathsf Y$ and $\mathsf X$. 
\begin{topic} {\bf Smooth functions on orbispaces}
  Let $U\subset X$ be an open subset of the $\mathfrak{T}$-orbispace $X$.
  A continuous function $g:U \rightarrow \R$ is called {\it smooth},
  if for every orbispace chart $(\tilde V,H,\upsilon)$  the composition
  $\upsilon^* (g) := g \circ \upsilon_{|\upsilon^{-1} (U)}$ is smooth. 
  The algebra of smooth functions $g:U \rightarrow \R$ will be denoted 
  by $\cC^\infty (U)$.
  The spaces $\cC^\infty (U)$ then form the sectional spaces 
  of a sheaf of algebras on $X$. We denote this sheaf by $\cC^\infty_X$ or
  briefly $\cC^\infty$ and call it the {\it sheaf of smooth functions} on $X$.
  By a {\it smooth map} between profinite dimensional orbispaces $X$ and $Y$ 
  we understand a continuous map $f:X\rightarrow Y$ such that
  that $f^* \cC^\infty_Y \subset \cC^\infty_X$.
  It is immediate to check that the $\mathfrak{T}$-orbispaces together with
  the smooth maps 
  between them form a category. 
  Moreover, it follows by a standard argument that the sheaf of 
  smooth functions on a $\mathfrak{T}$-orbispace is fine.

  Note that our definition of smooth maps is in correspondence with
  the smooth maps between orbit spaces in 
  \cite{Bie:SOSSEM,SjaLer:SSSR,Pfl:AGSSS}, 
  but that it is weaker than the notion of 
  smooth maps as defined in \cite{Sat:GNM,Rua:SGTO,CheRua:OGWT} 
  for the case of orbifolds.

  A particularly useful characterization of the smooth functions
  on a finite dimensional orbispace can be given as follows.
  Let $(\tilde U = G\times_H \tilde W,G ,\varrho)$ be a 
  twisted-linear orbispace chart for $X$ that means $H\subset G$ is  
  a closed subgroup and $\tilde W$ an open and 
  convex neighborhood of the origin of some $H$-representation space 
  $\mathfrak W$.
  Clearly, by slice theorem there exists an atlas for $X$ consisting of
  twisted-linear charts.  
  Choose a homogeneous Hilbert basis 
  $p= (p_1,\cdots,p_k)$ of the algebra $\mathcal P (\mathfrak W)^H$ of 
  $H$-invariant polynomials on $\mathfrak W$. 
  Since the Hilbert basis $p$ consists of $H$-invariant
  functions, the map
  \begin{displaymath}
    p_U: U \rightarrow \R^k, \: x \mapsto p(v) \quad 
    \text{with $v\in \tilde W$ such that $\varrho ([e,v]) =x$}
  \end{displaymath}
  is well-defined and continuous. Moreover, $p_U$ has the following
  properties:
  \begin{enumerate}[(1)]
  \item
    $p_U$ is a homeomorphism onto its image,
  \item 
    on every stratum of $U$, $p_U$ restricts to a diffeomorphism
    onto a smooth submanifold of $\R^k$,
  \item 
    the sheaf $\cC^\infty_U$ coincides with the pullback sheaf 
    $p_U^* \cC^\infty_{\R^k}$; this is a consequence of the theorem of 
    {\sc Schwarz} \cite{Schwa:SFIACLG}.
  \end{enumerate}
  In other words these properties mean that $p_U$ is a smooth chart 
  for the stratified space $X$ in the sense of
  \cite[Sec.~1.3]{Pfl:AGSSS}. From that one can derive the following
  result.
\end{topic}

\begin{proposition}
\label{PropSmoothFct}
  A continuous map $f:X \rightarrow X'$ between orbispaces is smooth,
  if and only if for all twisted linear charts
  $(\tilde U = G\times_H \tilde W,G ,\varrho)$ of $X$ and 
  $(\tilde U' = G'\times_{H'} \tilde W',G' ,\varrho')$ of $X'$ 
  such that
  $f(U) \subset U'$ there exists a smooth map 
  $\hat f_{UU'} :O \rightarrow \R^{k'}$ defined on an open 
  neighborhood $O \subset \R^k$ of $p_U (U)$ such that
  \begin{displaymath}
    \hat f_{UU'} \circ p_U = p'_{U'}\circ f_{|U}.
  \end{displaymath}
\end{proposition}

\begin{topic} {\bf Vector orbibundles }
  By a {\it vector orbibundle} we understand an orbispace $E$ which is 
  induced by an  {\it orbibundle functor} that means by an orbispace
  functor $\mathsf E$ having values in the category of vector bundles.
  We denote an orbibundle functor as follows:
  \begin{displaymath}
    \mathsf E :\clos{\cU}\rightarrow\mathfrak{VBdl}^{\text{\sf\tiny sym}}, 
    \quad 
    \begin{cases}
      u\mapsto (\tilde E_u,G_u),\\
      (v \rightarrow u) \mapsto
      (\psi_{vu},\iota_{vu}): (\tilde E_v,G_v)\rightarrow (\tilde E_u,G_u).
    \end{cases}
  \end{displaymath}
  A $\mathfrak{VBdl}$-orbispace chart for $E$ will be called an 
  {\it orbibundle chart}.   
  Similarly to the manifold case, a vector orbibundle gives rise to a base 
  orbispace and a canonical projection. Let us show this in more detail.
  Denote for every $u \in \clos{\cU}$ 
  by $\tilde U_u$ the base of the vector bundle $\tilde E_u$ and 
  by $\pi_u: \tilde E_u \rightarrow \tilde U_u$ the canonical
  projection. Moreover, let 
  $\varphi_{vu} : \tilde U_v \rightarrow \tilde U_u$  
  be the embedding on the level of base manifolds 
  induced by the morphism $\psi_{vu}$.
  Then 
  \begin{displaymath}
    \mathsf X: \clos{\cU} \rightarrow \mathfrak{Man}^{\text{\sf \tiny sym}},
    \quad 
    \begin{cases}
      u \mapsto (\tilde U_u, G_u),\\
      (v\rightarrow u) \mapsto (\varphi_{vu}, \iota_{vu}): 
      (\tilde U_v, G_v)\rightarrow (\tilde U_u, G_u) 
    \end{cases}
  \end{displaymath}
  is an orbispace functor. 
  The resulting orbispace $X$ is the {\it base orbispace} 
  of the vector orbibundle $E$.
  Clearly, every orbibundle chart $(E,G,\eta)$ of $E$ now induces
  an orbispace chart $(X,G,\varrho)$ on $X$ by the same procedure.  
  Note that even if $(E,G,\eta)$ is a reduced orbibundle chart, 
  $(X,G,\varrho)$ need not be reduced, in general.  
  Following {\sc Chen--Ruan} \cite{CheRua:OGWT} we say that $E$ is a 
  {\it good} or {\it reduced vector orbibundle}, if for every 
  reduced orbibundle chart
  $(E,G,\eta)$ of $E$ the induced chart $(X,G,\varrho)$ on the base
  is reduced as well.

  Next consider the canonical projections 
  $\pi_u : \tilde E_u \rightarrow \tilde U_u$, $u\in \clos{\cU}$. 
  Obviously, the $\pi_u$ induce a unique smooth map 
  $\pi: E \rightarrow X$ called {\it projection} such that
  \begin{equation}
  \label{Eq2.2}
    \pi \circ \eta_u  =  \varrho_u \circ \pi_u  \quad
    \text{for all $u\in \clos{\cU}$}.
  \end{equation}
  Analogously like for vector bundles one  defines a {\it section} of $E$  
  as a continuous map $s: X\rightarrow E$ such that $\pi\circ s =\id_X$.
  We denote the space of continuous (resp.~smooth) sections of $E$ by
  $\Gamma (E)$ (resp.~$\Gamma^\infty (E)$).
  But unlike in the case of vector bundles, an orbibundle
  $E\rightarrow X$ is in general not locally trivial over the base,
  which implies in particular that the space of continuous resp.~smooth
  sections need not be linear. In the following, we will construct for every 
  vector orbibundle a subspace 
  $\Gamma_{\text{\sf \tiny str}}^\infty (E) \subset \Gamma^\infty (E)$ which
  is a $\cC^\infty (X)$-module in a natural way.
  The elements of $\Gamma_{\text{\sf \tiny str}}^\infty (E)$ will be called
  {\it smooth stratified sections} of $E$. 
  To define $\Gamma_{\text{\sf \tiny str}}^\infty (E)$ 
  let $(\tilde E,G,\eta)$ be an 
  orbibundle chart for $E$ and $(\tilde U ,G ,\varrho)$ the induced orbispace
  chart for the base. For every point $\tilde x\in \tilde U$ let 
  $\tilde E_{\tilde x}^{G_{\tilde x}}$ be the (linear) subspace  
  of $G_{\tilde x}$-invariant elements of the fiber $\tilde E_{\tilde x}$.
  Then for every closed subgroup $H\subset G$  
  \begin{displaymath}
    \tilde E_{(H)} := \bigcup_{\tilde x \in \tilde U\atop (G_{\tilde x})=(H)}
    \tilde E_{\tilde x}^ {G_{\tilde x}}
  \end{displaymath}
  is a smooth vector bundle over the stratum $\tilde U_{(H)}$ and
  $\tilde E_{(H)} /G$  a smooth vector bundle over $M_{(H)}/G$. Moreover,
  one concludes easily by the slice theorem that $\tilde E_{(H)}$ can be 
  identified with the pullback bundle of 
  $\tilde E_{(H)} /G \rightarrow \tilde U_{(H)} /G$ by the canonical
  projection $\tilde U_{(H)} \rightarrow \tilde U_{(H)} /G$.
  Now, the union
  \begin{displaymath}
    E_u^{\text{\sf\tiny str}}:=\bigcup_{(H)\subset G}\tilde E_{(H)}/G    
  \end{displaymath}
  is a (in general not closed) subspace of $\tilde E/G$ which carries
  a canonical stratification given by 
  the set germs of the vector bundles $\tilde E_{(H)}/G$.
  The only nontrivial part in the proof of this is to show that locally, 
  the condition of frontier (DEC\ref{IteDecSp2}) is satisfied.  
  To this end it suffices to prove
  that for all orbit types $(K) \subsetneq (H)$  
  and every point $\tilde x\in \tilde U_{(H)} \cap \overline{\tilde U_{(K)}}$ 
  the fiber 
  $\tilde E_{\tilde x}^{G_{\tilde x}}$ is contained in the closure
  of $\tilde E_{(K)}$. Let us show this.
  By the slice theorem we can assume after possibly passing to conjugate 
  subgroups that
  $G_{\tilde x} = H$, $K\subset H$ and that there exists a sequence of 
  points $\tilde x_n \in \tilde U_{K}$ converging to $\tilde x$.
  By passing to an appropriate subsequence of $(\tilde x_n)$ we can
  achieve that the
  sequence of fibers $\tilde E_{\tilde x_n}^K$ converges in the bundle
  of Grassmannians.
  By  $K \subset H$ one concludes that
  \begin{displaymath}
    \tilde E_{\tilde x}^H \subset \lim_{n\rightarrow \infty} \,
    \tilde E_{\tilde x_n}^K,
  \end{displaymath}
  whence the condition of frontier holds true.
 
  Next, consider an open embedding 
  $(\psi_{vu},\iota_{vu}): (\tilde E_v,G_v,\eta_v)\rightarrow 
  (\tilde E_u,G_u,\eta_u)$ between orbibundle charts of $\mathsf E$. 
  Then, the induced map between the 
  orbit spaces restricts to a strata preserving open embedding  
  \begin{displaymath}
    \overline{\psi}_{vu}^{ \, \text{\sf\tiny str}} : \: 
    \tilde E_v^{\text{\sf\tiny str}} \rightarrow 
    \tilde E_u^{\text{\sf\tiny str}}.
  \end{displaymath}
  Restricted to a stratum, $\overline{\psi}_{vu}^{\,\text{\sf\tiny str}}$ 
  is a smooth vector bundle isomorphism onto an open subbundle of the 
  image stratum.  Hence, the union
  \begin{displaymath}
    E^{\text{\sf\tiny str}} = \bigcup_{u \in \clos{\cU}} \,
    \overline{\eta}_u (E_u^{\text{\sf\tiny str}})  \subset E
  \end{displaymath}
  carries a uniquely defined structure of a stratified space
  such that every one of the topological embeddings
  $\overline{\eta}_u : \tilde E_u/G \rightarrow E$ is an
  isomorphism of stratified  spaces 
  from $E_u^{\text{\sf\tiny str}}$ onto an open subset of
  $E^{\text{\sf\tiny str}}$.
  We will say that $E^{\text{\sf\tiny str}}$ is the 
  {\it stratified vector bundle associated} to the  vector orbibundle
  $E$. A smooth section $s:X \rightarrow E$ with
  image in $E^{\text{\sf\tiny str}}$ now will be called a 
  {\it smooth stratified section}, if it satisfies the following
  smooth vertical extension property:
  \begin{enumerate}[(SVX)]
  \item 
    For sufficiently small $\varepsilon >0$ the vertical extension
    \begin{displaymath}
      \mathsf V_s : E \times (-\varepsilon , \varepsilon) \rightarrow E, \quad
      (v,t) \mapsto v + t s (\pi (v)) 
    \end{displaymath}
    is smooth.
  \end{enumerate}
  By construction of $E^{\text{\sf\tiny str}}$, the vertical extension
  is well-defined and continuous. Clearly, whether $\mathsf V_s$ is smooth,
  depends only on the (maximal) orbibundle atlas of $E$ and not
  on the particular defining orbibundle functor $\mathsf E$.
  The space of smooth stratified sections will be denoted by 
  $\Gamma^\infty_{\text{\sf\tiny str}} (E) $ or 
  $\Gamma^\infty( E^{\text{\sf\tiny str}})$.
  \begin{proposition}
  \label{ProRedBdl}
    Let $E\rightarrow X$ be a vector orbibundle.
    Then the following relations are equivalent:
    \begin{enumerate}[(1)]
      \item
      \label{IteRed1}
        $E$ is a reduced vector orbibundle.
      \item 
      \label{IteRed2} 
        $E^{\text{\sf\tiny str}}$ is dense in $E$.
      \item
      \label{IteRed3} 
        The projection 
        $E^\circ_{|X^\circ} := 
        E^\circ \cap \pi^{-1} (X^\circ)\rightarrow X^\circ$ 
        is a smooth vector bundle.
    \end{enumerate}
  \end{proposition}
  \begin{proof}
   Let us first show that (\ref{IteRed1}) implies (\ref{IteRed3}).
   Let $E\rightarrow X$ be reduced and $x\in X^\circ$ be a point.
   Choose a slice orbibundle chart $(\tilde E\rightarrow \tilde U,G,\eta)$ 
   around $0_x\in E$, and $\tilde x \in \tilde U$ with $\varrho (\tilde x) =x$.
   By restriction to an appropriate open subbundle of $\tilde E$,
   we can achieve that $\tilde U /G$ lies in the regular part of $X$.
   Moreover, after passing to the reduced orbibundle chart, we can assume that
   $G$ acts effectively on $\tilde E$. Hence, by assumption, $G$ acts
   effectively on $\tilde U$. Since $(\tilde E,G,\eta)$ is a slice 
   for the orbibundle germ at $0_{\tilde x} \in \tilde E$, the orbichart
   $(\tilde U, G,\varrho)$ is a slice for the orbispace germ 
   at $\tilde x$. Thus $G_{\tilde y} = G$ for all $\tilde y \in \tilde U$.
   But $G$ acts effectively on $\tilde U$, so $G = \{e\}$.
   From this one concludes  that  
   $E_{|U} := \pi^{-1} (U) = \tilde E$, 
   hence $E_{|U} \subset E^\circ$. By definition of $E^\circ_{|X^\circ}$,
   (\ref{IteRed3}) follows.

   Clearly, $E^\circ_{|X^\circ} \rightarrow X^\circ$ is a vector bundle, 
   if and only if 
   $E^{\text{\sf\tiny str}}\cap \pi^{-1} (X^\circ) =E^\circ_{|X^\circ}$. 
   Hence (\ref{IteRed2}) and (\ref{IteRed3}) are equivalent. 

   For the proof of the implication 
   (\ref{IteRed3}) $\Rightarrow$ (\ref{IteRed1}) let $(\tilde E,G,\varrho)$ be
   reduced and $v\in E^\circ_{|X^\circ} \cap \eta (\tilde E)$.
   Then $G_{\pi (v)} =G_v$ by definition of $E^{\text{\sf\tiny str}}$. Hence
   $\bigcap_{g\in G} G_{\pi (v)} = \bigcap_{g\in G} G_v =\{ e\}$,
   so \ref{ExOrbiType} (\ref{StraOrbiIte3})
   entails  that $G$ acts effectively on $\tilde U$.
  \end{proof}
\end{topic}

\begin{example}
  Let $X$ be an orbispace. Then the {\it tangent orbibundle functor} 
  $T\mathsf X:\clos{\cU} \rightarrow \mathfrak{VBdl}^{\text{\sf \tiny sym}}$ 
  is defined to be the functor which associates to every orbispace chart 
  $(\tilde U,G,\varrho)$ of $\mathsf X$ the object $(T \tilde U,G)$ and to 
  every morphism 
  $(\varphi,\iota)= \mathsf X_ {VU}:
  (\tilde V,H,\upsilon)\rightarrow (\tilde U,G,\varrho)$ 
  the morphism
  $(T\varphi,\iota):(T \tilde V,H)\rightarrow (T \tilde U,G)$.
  The (finite dimensional) orbibundle defined by $T\mathsf X$ will be called
  the {\it tangent orbibundle} of $X$ and will be denoted by $TX$.
  Similarly, one defines the {\it cotangent orbibundle} $T^*X$.
  Note that both the tangent and cotangent orbibundles are good orbibundles.
  
  More generally, if $\mathsf{F}$ is a functor on the category of 
  (finite dimensional) real or complex vector spaces and 
  $\mathsf E:\clos{\cU}\rightarrow \mathfrak{VBdl}^{\text{\sf \tiny sym}}$ 
  an orbibundle functor, then
  the fiberwise application of $\mathsf{F}$ to every one of the objects
  $\mathsf E (u)$ leads to a new vector orbibundle functor 
  denoted by $\mathsf{FE}$. Generalizing this even further to
  covariant and contravariant functors in multiple arguments it is then
  clear what to understand by the direct sum, the tensor product and so on
  of vector orbibundles over a common base orbispace $X$. In the remainder of 
  this work we will use such constructions of vector orbibundles
  without further explanation.
\end{example}
\begin{theorem}
\label{ThmGSec}
  Let $E$ be an orbibundle over an orbispace $X$.
  Then the space $\Gamma^\infty_{\text{\sf\tiny str}} (E) $ of smooth 
  stratified sections carries a natural structure of a 
  $\cC^\infty (X)$-module. Moreover, if $\cU$ is an open covering of $X$ and 
  $\mathsf E:\clos{\cU}\rightarrow \mathfrak{VBdl}^{\text{\sf\tiny sym}}$
  an orbibundle functor of $E$ inducing the orbispace functor $\mathsf X$  
  on the base, then a continuous section $s:X \rightarrow E$
  is a smooth stratified section, if and only if it is is a good section.
  $s$ being a good section  hereby means that  there exists a family 
  $(s_{\tilde U})_{U\in\clos{\cU}}$ of smooth sections 
  $s_{\tilde U} : \tilde U \rightarrow \tilde E_{U}$ such that the 
  following conditions hold true:
  \begin{enumerate}[(GSEC1)]
  \item 
  \label{GSEC1}
    For every orbispace chart $(\tilde U,G,\varrho)$ of $\mathsf X$ 
    the section  $s_{\tilde U}$ is $G$-equivariant.
  \item 
  \label{GSEC2}
    If $(\varphi_{VU},\iota_{VU}) = \mathsf X_{VU}: 
    (\tilde V,H,\upsilon)\rightarrow (\tilde U,G,\varrho)$ 
    is a morphism and $(\psi_{VU},\iota_{VU}) = \mathsf E_{VU}$ 
    the corresponding morphism between the vector bundles 
    $(\tilde E_{V},H)$ and $(\tilde E_{U},G)$, then
    \begin{equation}
      s_{\tilde U} \circ \varphi_{VU} = \psi_{VU} \circ s_{\tilde V}. 
    \end{equation}
  \item
  \label{GSEC3}
    For every $(\tilde U,G,\varrho)$ the following relation holds true:
    \begin{equation}
      \eta_{U}\circ s_{\tilde U} = s\circ \varrho .
    \end{equation}
  \end{enumerate}
  If $s$ is a smooth stratified section, then the family $(s_{\tilde U})$ 
  satisfying (GSEC\ref{GSEC1}) to (GSEC\ref{GSEC3}) is uniquely determined.
\end{theorem}
\begin{remark}
  The notion of {\it good maps} between orbifolds has been introduced
  by {\sc Chen--Ruan} \cite{CheRua:OGWT} in their work on orbifold
  Gromov--Witten theory. The essential feature hereby
  is that the pull-back of a vector orbibundle by a 
  good map is a well-defined concept, whereas the pull-back 
  orbibundle of an arbitrary smooth map does
  in general not exist.
  Moreover, good maps between orbifolds correspond to the morphisms
  of orbifolds as defined in the groupoid approach to orbifolds. 
  See {\sc Moerdijk} \cite{Moe:OGI} for more on this. 
\end{remark}
\begin{proof}
  Clearly, the second claim implies the first, so we 
  only show that $s$ is a smooth stratified section if and only if it is a
  good section.
  The existence of a family $(s_{\tilde U})$  satisfying
  (GSEC\ref{GSEC1}) to (GSEC\ref{GSEC3}) is obviously sufficient 
  for $s$ to be a smooth stratified map. Hence it 
  remains to prove that the existence of such a family $(s_{\tilde U})$ is 
  also necessary. For simplicity we assume that $\cU$ consists only of one 
  connected open set $U$ or, in other words, that $E$ is the orbit space of
  the orbibundle chart $(\tilde E,G,\eta) = \mathsf E (U)$.
  The general case can easily be deduced from this particular situation.
  Under the assumption made for $E$, let $s$ be a smooth 
  stratified section
  $s: \tilde U/G \rightarrow \tilde E/G$.
  Now, given $f\in \cC^\infty (\tilde E/G)$ we define a function
  $\delta_s f\in \cC^\infty (\tilde E/G)$  as follows:
  \begin{equation}
  \label{EqDefDer}
     \delta_s f (v) = \frac{d}{dt} \left.
     f \big( \mathsf V_s (v ,t )\big) \right|_{t=0}
     \quad\text{for all $v\in \tilde E/G$},
  \end{equation}
  where $\mathsf V_s$ is the smooth vertical extension of $s$.
  By construction $\delta_s$ is a derivation on $\cC^\infty (\tilde E/G)$.
  Hence, according to the Smooth Lifting Theorem of 
  {\sc Schwarz} \cite[Thm.~0.2]{Schwa:LSHOS}, there exists a 
  $G$-invariant smooth vector field  
  $\xi : \tilde E \rightarrow T \tilde E$ such that 
  \begin{displaymath}
    \xi (f\circ \eta) = \delta_s f \quad 
    \text{for all $f\in \cC^\infty (\tilde E/G)$}.
  \end{displaymath}
  Obviously, $\xi$ is a vertical vector field, since the restriction of
  $\delta_s$ to $E^\circ_{|X^\circ}$ is vertical. 
  One concludes $s \circ \varrho = \xi_{|\tilde U}$, where 
  $\tilde U$ has been identified with the zero section of $\tilde E$. 
  Let us put $s_{\tilde U} := \xi_{|\tilde U}$.
  Then, $s_{\tilde U}$ is a smooth $G$-invariant section of $\tilde E$
  and satisfies
  \begin{equation}
  \label{EqInvSec}
     \eta \circ s_{\tilde U} = s \circ \varrho.
  \end{equation}
  Thus (GSEC\ref{GSEC1}) and (GSEC\ref{GSEC3}) hold true.
  
  Next let us show that the $G$-invariance and 
  Eq.~(\ref{EqInvSec}) uniquely determine $s_{\tilde U}$. 
  To this end check first that
  $s_{\tilde U} (\tilde x) \in \tilde E^{G_{\tilde x}}_{\tilde x}$ for
  all $\tilde x \in \tilde U$. 
  Second recall that for every $x\in U$ the fiber 
  $E^{\text{\sf \tiny str}}_x$ 
  coincides naturally with $\tilde E^{G_{\tilde x}}_{\tilde x}$, where
  $\tilde x \in \varrho^{-1} (x)$. By Eq.~(\ref{EqInvSec})
  this entails that $s_{\tilde U}$ is uniquely determined. 
  
  Finally, if $\cU$ is an arbitrary open covering of $X$,
  axiom (GSEC\ref{GSEC2}) 
  follows immediately from the uniqueness of the sections
  $s_{\tilde U}$, since for $V, U \in \cU$ with $V\subset U$
  the composition
  $\psi^{-1}_{VU} \circ s_{\tilde U} \circ \varphi_{VU}$ 
  is also a $G$-equivariant section over $\tilde V$ 
  satisfying (GSEC\ref{GSEC3}), hence it must coincide with $s_{\tilde V}$. 
\end{proof}
\begin{remark}
\label{RemSecRedOrbiBdl}
 According to the theorem one can identify a smooth stratified section of 
 a reduced vector orbibundle with a family $(s_{\tilde U})_{\tilde U\in\cU}$ 
 having properties (GSEC\ref{GSEC1}) to (GSEC\ref{GSEC3}), and every family 
 $(s_{\tilde U})_{\tilde U\in\cU}$ which fulfills (GSEC\ref{GSEC1}) and 
 (GSEC\ref{GSEC2}) gives rise to a unique smooth stratified section such that 
 also (GSEC\ref{GSEC3}) holds true.
 In the rest of this work we will very often make use of these canonical 
 identifications. For example we denote vector fields 
 $\xi :X \rightarrow TX$ briefly by $(\xi_{\tilde U})$ and
 assume from now on that the index $\tilde U$ runs through the domains of 
 the orbispace charts of the defining orbifold functor $\mathsf X$.
 Likewise, we denote differential forms on $X$, tensor fields and so on.   
\end{remark}


%
%
\section{Symplectic orbispaces}
\label{SecSympOrbi}
\begin{topic}
  Let $X$ be an orbispace, and $\cU,\mathsf X$ like before.
  By a {\it riemannian metric} (resp.~{\it symplectic form})  on $X$ 
  we understand a family of 
  $G$-invariant riemannian metrics $g_{\tilde U}$ (resp.~symplectic forms
  $\omega_{\tilde U}$) on $\tilde U$, where $(\tilde U,G,\varrho)$ runs
  through the charts of $\mathsf X$,
  such that for every every morphism 
  $(\varphi,\iota) := \mathsf X_{VU}: (\tilde V ,H,\upsilon) \rightarrow 
  (\tilde U,G,\varrho)$ between two orbispace charts the relation 
  \begin{align}
  \label{CompRie}
     \varphi^* g_{\tilde U}  & = g_{\tilde V} \quad \text{resp.}\\ 
  \label{CompSymp}
     \varphi^*  \omega_{\tilde U}  & = \omega_{\tilde V}
  \end{align}
  is satisfied. We will denote such a riemannian metric (resp.~symplectic
  form) by $(g_{\tilde U})$ (resp.~$(\omega_{\tilde U})$).
  An orbispace with a riemannian metric 
  $(g_{\tilde U})$ (resp.~symplectic form
  $(\omega_{\tilde U})$) 
  will be called a {\it riemannian} (resp.~{\it symplectic}) 
  {\it orbispace}; likewise one defines {\it riemannian} and 
  {\it symplectic orbifolds}.
  Note that by Thm.~\ref{ThmGSec}, $(g_{\tilde U})$ 
  (resp.~$(\omega_{\tilde U})$) corresponds to a smooth stratified section
  $g\in \Gamma^\infty_{\text{\sf \tiny str}} (T^*X\otimes T^*X)$ 
  (resp.~$\omega\in \Gamma^\infty_{\text{\sf \tiny str}} (T^*X\otimes T^*X)$).

  Since for every orbispace chart $(\tilde U, G, \varrho)$ there 
  exists a $G$-invariant riemannian metric on $\tilde U$ and because the sheaf
  $\cC^\infty_X$ is fine, it is easy to construct
  a riemannian metric for $X$.

  Like in the manifold case, natural examples of symplectic orbispaces
  are given by cotangent bundles. 
  To see this, let $T^*X$ be the cotangent orbibundle of $(X,\mathsf X)$ 
  and consider the orbispace chart $(T^*\tilde U, G, T^*\varrho)$ induced 
  by $(\tilde U, G, \varrho)$. Then $T^*\tilde U$ carries a canonical
  symplectic form $\omega_{T^*\tilde U}$ and this symplectic form  is 
  invariant with respect to the lifted $G$-action. Moreover, if 
  $(\varphi,\iota) : (\tilde V, H, \upsilon)\rightarrow (\tilde U, G, \varrho)$
  is a morphism and 
  $(T^*\varphi,\iota) =({\varphi^{-1}}^*,\iota):
  (T^*\tilde V, H, T^*\upsilon)\rightarrow (T^*\tilde U, G, T^*\varrho)$ 
  the induced morphism of orbispace charts of $T^*X$, then 
  $(T^*\varphi)^* \omega_{T^*\tilde U}= \omega_{T^*\tilde V}$  ,
  hence the $\omega_{T^*\tilde U}$ define a symplectic form on $T^*X$.
\end{topic}
\begin{example}
\label{ExSympCone}
  As a specific example of a symplectic orbifold consider the cotangent
  orbibundle of the real half line $Y =[0,\infty)$.
  A global orbifold chart for $Y$ is given by $\R$ with the 
  $\Z_2$-action such that the nonzero element acts by inversion. 
  Therefore, $T^*Y$ is the quotient $\R^2/ \Z_2$, where 
  the nonzero element of $\Z_2$ acts again by inversion.
  A Hilbert basis of the $\Z_2$-invariant polynomials on $\R^2$ is given by
  the polynomials $p^2+q^2$, $p^2-q^2$ and $2pq$, where $(p,q)$ are the
  coordinates of an element of $\R^2$. Now,
  \begin{displaymath}
    (p^2+q^2)^2 = (p^2-q^2)^2 + (2pq)^2,
  \end{displaymath}
  hence the orbifold $\R^2/\Z_2$ is diffeomorphic to the
  standard cone  $\{ (x_1,x_2,x_3) \in \R^3 \mid x_1^2+x_2^2 =x_3^2\}$.
  Moreover, the symplectic orbifold $\R^2/\Z_2$ has a natural 
  stratification by two symplectic strata, where the top stratum is given
  by $\dot{\R}^2/\Z_2$ with $\dot{\R}^2 = \R^2 \setminus \{ 0\}$
  and the second stratum is given by $\{0\}$ or in other words by
  the cusp of the cone. 
\end{example}
\begin{proposition}
  Let $X$ be a symplectic orbispace. Then every stratum of the orbispace
  stratification carries in a canonical way the structure of a Poisson
  manifold. Moreover, if $X$ is an orbifold, the strata are symplectic.
\end{proposition}
\begin{proof}
  We show the claim for the case, where the orbispace is given by the 
  orbit space of a symplectic $G$-action on a symplectic manifold $M$. 
  Clearly, this suffices to prove the proposition, since the claimed
  property of $X$ is essentially a local statement.
  So let us assume that $X=M/G$. Then it is well-known that for every
  orbit type $(H)$ the manifold $M_H$ of points of $M$ with isotropy
  group equal to $H$ inherits from $M$ a symplectic structure 
  \cite[Prop.~27.5]{GuiSte:STP}.
  Moreover, the canonical projection $\pi_{H}: M_H \rightarrow M_{(H)}/G$ onto
  the stratum  $M_{(H)} /G$ is a principal fiber bundle with typical
  fiber $N_G (H)/H$, where $N_G(H)$ is the normalizer of $H$ in $G$.
  Now, given two functions $f,g \in \cC^\infty (M_{(H)}/G)$ the Poisson
  bracket $\{ f\circ \pi_H,g\circ \pi_H\}$ with respect to the canonical 
  symplectic structure on $M_H$ is $N_G(H)$-invariant, hence there
  exists a unique $\{ f,g\}_H \in \cC^\infty (M_{(H)}/G)$ such that
  \begin{displaymath}
    \{ f,g \}_H \circ \pi_H = \{ f\circ \pi_H,g\circ \pi_H \} .
  \end{displaymath}
  Clearly, $\{\cdot,\cdot\}_H$ is antisymmetric and satisfies the Jacobi 
  identity, hence is a Poisson bracket on $\cC^\infty (M_{(H)}/G)$.
  Thus, $M_{(H)}/G$ carries the structure of a Poisson manifold 
  and this Poisson structure is natural in the sense that it is
  invariant under equivariant symplectic diffeomeorphisms of $M$.
  
  Under the assumption that the symmetry group $G$ is finite
  the zero map $M\rightarrow \{0\}$ provides a momentum map for the 
  symplectic $G$-action, so by {\sc Sjamaar--Lerman}
  \cite[Thm.~2.1]{SjaLer:SSSR} the strata  $M_{(H)}/G$ are symplectic
  in this case. This proves the proposition.
\end{proof}
\begin{topic}
  A family $(\nabla_{\! \tilde U})$ of $G$-invariant 
  (affine) connections $\nabla_{\! \tilde U}$ defined on 
  $\Gamma^\infty (T\tilde U)$ is called
  a {\it connection} on $X$, if for every vector field $(\xi_{\tilde U})$
  on $X$ and every morphism 
  $(\varphi,\iota) : (\tilde V,H,\upsilon) \rightarrow (\tilde U,G,\varrho)$ 
  between charts of $\cU$ the compatibility relation
  \begin{equation}
      \varphi^* (\nabla_{\!\tilde U} \xi_{\tilde U}) = 
      \nabla_{\!\tilde V} \xi_{\tilde V}
  \end{equation}
  holds true. Note that every
  connection $(\nabla_{\!\tilde U})$ on $X$ gives rise to a
  {\it covariant derivative}, i.e.~a linear map
  $\nabla : \Gamma^\infty_{\text{\sf \tiny str}} (TX)\rightarrow 
  \Gamma^\infty_{\text{\sf \tiny str}} (T^*X \otimes TX)$
  such that
  \begin{equation}
    \nabla (f \xi) = df \otimes \xi + f \, \nabla \xi \quad 
    \text{for all $f\in \cC^\infty (X) $ and $\xi \in 
    \Gamma^\infty_{\text{\sf \tiny str}} (TX)$}.
  \end{equation}

  If $(g_{\tilde U})$ is a riemannian metric on $X$, then the   
  family $(\nabla_{\!\tilde U}^{^{\text{\tiny LC}}})$, 
  which associates to every $\tilde U$ the 
  Levi--Civita connection with respect to $g_{\tilde U}$, provides a 
  torsionfree connection on $X$. Obviously, 
  $(\nabla_{\!\tilde U}^{^{\text{\tiny LC}}})$
  leaves the riemannian metric $(g_{\tilde U})$ invariant and will be 
  called the {\it Levi--Civita connection} of $(g_{\tilde U})$.
  In case $(\omega_{\tilde U})$ is a 
  symplectic form  on $X$, a connection $(\nabla_{\!\tilde U})$ is called 
  {\it symplectic}, if $\nabla_{\!\tilde U} \omega_{\tilde U} =0$
  holds for all $\tilde U$.
  
  More generally, let us assume now that $E\rightarrow X$ is a reduced vector 
  orbibundle, where the typical fiber $V$ is a profinite dimensional 
  vector space. 
  By a {\it connection} on $E$ we then understand a linear map
  $D :\Gamma^\infty_{\text{\sf \tiny str}} (\Lambda^\bullet X \otimes E) 
  \rightarrow \Gamma^\infty_{\text{\sf \tiny str}} 
  (\Lambda^\bullet X \otimes E )$ of antisymmetric degree $1$ such that
  \begin{equation}
    D (\alpha \wedge s) = d\alpha \wedge s + (-1)^k \, \alpha \wedge Ds \quad
    \text{for all $\alpha\in \Gamma^\infty (\Lambda^k X)$ and 
    $s \in \Gamma^\infty_{\text{\sf \tiny str}} (E)$}.
  \end{equation}
  Given a  Satake atlas $\cU$ for $X$ and a bundle atlas 
  $\big( (E_{\tilde U}, G,\eta_{\tilde U})\big)_{\tilde U\in\cU}$ over
  $\cU$, Thm.~\ref{ThmGSec} entails that a connection can be 
  regarded as a family $(D_{\tilde U})$ of connections
  $D_{\tilde U}:\Gamma^\infty (\Lambda^\bullet\tilde U\otimes E_{\tilde U})
  \rightarrow \Gamma^\infty(\Lambda^\bullet\tilde U\otimes E_{\tilde U})$ 
  such that for every smooth section $s =(s_{\tilde U})$ one has 
  \begin{equation}
    (Ds)_{\tilde U} = D_{\tilde U} s_{\tilde U} \quad 
    \text{for all $\tilde U \in \cU$}.
  \end{equation}
  The {\it curvature} of a connection $D$ is the two-form
  $R\in\Gamma^\infty_{\text{\sf \tiny str}}  (\Lambda^2 X\otimes \End(E))$ with
  \begin{equation}
    R (\xi,\zeta) \, s = [D_\xi , D_\zeta]\,s - 
    D_{[\xi,\zeta]}\, s
    \quad \text{for all $\xi,\zeta \in \Gamma^\infty_{\text{\sf \tiny str}}(TX)$
    and $s\in \Gamma^\infty_{\text{\sf \tiny str}} (E)$.}
  \end{equation}
  Obviously, $R =(R_{\tilde U})$, where $R_{\tilde U}$ is
  the curvature of $D_{\tilde U}$. 

\end{topic}
\begin{proposition}
  For every symplectic orbispace there exists a torsionfree symplectic
  connection. 
\end{proposition}
\begin{proof}
  First  fix a riemannian metric $(g_{\tilde U})$ on $X$ and
  use the corresponding Levi--Civita connection
  $(\nabla_{\!\tilde U}^{^{\text{\tiny LC}}})$ 
  to define a contravariant $3$-tensor field 
  $(\Delta'_{\tilde U})$ on $TX$:
  \begin{equation}
    \Delta'_{\tilde U} (\xi_1,\xi_2,\xi_3) := \frac 13 
    \left( \nabla_{\!\tilde U}^{^{\text{\tiny LC}}} 
    \omega_{\tilde U} (\xi_3,\xi_1,\xi_2) +
    \nabla_{\!\tilde U}^{^{\text{\tiny LC}}} 
    \omega_{\tilde U} (\xi_2,\xi_1,\xi_3) \right), 
    \quad \xi_1,\xi_2,\xi_3 \in T_{\tilde x}\tilde U,\:\tilde x\in \tilde U .
  \end{equation}
  Note that $(\Delta'_{\tilde U})$ is symmetric in the last two
  variables. 
  Next lift the first variable of $(\Delta'_{\tilde U})$ with the help of 
  $(\omega_{\tilde U})$ and denote the resulting tensor field 
  by $(\Delta_{\tilde U})$, that means the equality
  $\omega_{\tilde U} (\:\cdot\:,\Delta_{\tilde U}) =\Delta'_{\tilde U}$ is
  satisfied over each $\tilde U$.
  Then by construction, the connection $(\nabla_{\!\tilde U})$ defined by
  \begin{equation}
    \nabla_{\!\tilde U} = \nabla_{\!\tilde U}^{^{\text{\tiny LC}}}
    + \Delta_{\tilde U}
  \end{equation}
  consists of $G$-invariant and torsionfree local connections. Moreover,  
  it is also clear by construction  that these connections satisfy the 
  compatibility condition $\varphi^* \nabla_{\tilde U} = \nabla_{\tilde V}$
  for every morphism $(\varphi,\iota)$ like above. 
  Finally, $(\nabla_{\!\tilde U})$ is symplectic 
  by the following computation:
  \begin{equation}
    \begin{split}
      \nabla_{\!\tilde U} \omega_{\tilde U} (\xi_1,\xi_2,\xi_3)  =\, & 
      \nabla_{\!\tilde U}^{^{\text{\tiny LC}}}
      \omega_{\tilde U} (\xi_1,\xi_2,\xi_3) -
      \omega_{\tilde U}
      (\nabla_{\tilde U}^{^{\text{\tiny LC}}} (\xi_1,\xi_2), \xi_3) -
      \omega_{\tilde U} (\xi_2, \nabla_{\!\tilde U}^{^{\text{\tiny LC}}}
      (\xi_1,\xi_3)) \\
      =\, & \nabla_{\!\tilde U}^{^{\text{\tiny LC}}}
      \omega_{\tilde U} (\xi_1,\xi_2,\xi_3) -
      \left( \Delta'_{\tilde U} (\xi_2,\xi_1,\xi_3) - 
      \Delta'_{\tilde U} (\xi_3,\xi_1,\xi_2) \right) \\
      =\, & \nabla_{\!\tilde U}^{^{\text{\tiny LC}}}
      \omega_{\tilde U} (\xi_1,\xi_2,\xi_3) -
      \frac 13\Big( \nabla_{\!\tilde U}^{^{\text{\tiny LC}}}
      \omega_{\tilde U}  (\xi_3,\xi_2,\xi_1)
      + \nabla_{\!\tilde U}^{^{\text{\tiny LC}}}
      \omega_{\tilde U}  (\xi_1,\xi_2,\xi_3) -\\
      & \hspace{35mm} -
      \nabla_{\!\tilde U}^{^{\text{\tiny LC}}} 
      \omega_{\tilde U}  (\xi_2,\xi_3,\xi_1) -
      \nabla_{\!\tilde U}^{^{\text{\tiny LC}}} 
      \omega_{\tilde U}  (\xi_1,\xi_3,\xi_2) \Big) \\
      =\, & \frac 13 \Big(
      \nabla_{\!\tilde U}^{^{\text{\tiny LC}}}
      \omega_{\tilde U} (\xi_1,\xi_2,\xi_3) +
      \nabla_{\!\tilde U}^{^{\text{\tiny LC}}}
      \omega_{\tilde U} (\xi_2,\xi_3,\xi_1) +
      \nabla_{\!\tilde U}^{^{\text{\tiny LC}}}
      \omega_{\tilde U} (\xi_3,\xi_1,\xi_2) \Big)\\
       = \, & d \omega_{\tilde U} (\xi_1,\xi_3,\xi_2) =  0.
    \end{split}
  \end{equation}
\end{proof}
\begin{topic}
  Given a symplectic form $(\omega_{\tilde U})$ on $X$ one can define a 
  natural Poisson bracket on the algebra $\cC^\infty (X)$ as follows.
  For every point $x\in X$ choose an orbispace chart 
  $(\tilde U,G,\varrho)$ around $x$, let $\tilde x \in \tilde U$ 
  be a point with $\varrho (\tilde x) =x$ and denote by 
  $\{\cdot , \cdot \}_{\tilde U}$ the Poisson bracket on 
  $\cC^\infty (\tilde U)$. Then define
  \begin{equation}
    \{ f, g\} (x) := \{ f\circ \varrho , g\circ \varrho \}_{\tilde U}
    (\tilde x) \quad \text{for $f,g \in \cC^\infty (X)$}.
  \end{equation}
  By the compatibility relation (\ref{CompSymp}), the value
  $\{f,g\} (x)$ is independent of the special choice of the chart 
  $(\tilde U,G,\varrho)$, so $\{ f ,g \}\in \cC^\infty (X)$ is well-defined.
  Using the corresponding properties of the Poisson brackets 
  $\{\cdot , \cdot \}_{\tilde U}$ one now checks immediately that 
  $\{ \cdot ,\cdot \}$ is antisymmetric in its
  arguments and satisfies the Jacobi identity, hence $\{ \cdot ,\cdot \}$
  is a Poisson bracket on $\cC^\infty (X)$.
  Note that the symplectic form $(\omega_{\tilde U})$ also gives rise
  to the {\it Poisson bivector field} $\Pi =(\Pi_{\tilde U})$ on $X$, where
  $\Pi_{\tilde U}$ is the Poisson bivector field on $\tilde U$ corresponding
  to $\omega_{\tilde U}$.
    
  The well-known definition of a formal deformation quantization of a
  symplectic manifold by 
  {\sc Bayen--Flato--Lichnerowicz--Sternheimer}  \cite{BayFlaFroLicSte:DTQ}
  can be easily extended to the the orbispace arena. 
  Let us provide the details.
  Consider the space $\cC^\infty (X)[[\lambda]]$
  of formal power series in the variable $\lambda$ and with coefficients
  in $\cC^\infty (X)$. A $\C[[\lambda]]$-bilinear associative product
  \begin{displaymath}
    \star: \cC^\infty (X)[[\lambda]] \times \cC^\infty (X)[[\lambda]]
    \rightarrow \cC^\infty (X)[[\lambda]]
  \end{displaymath}
  is called a {\it formal deformation quantization} of $\cC^\infty (X)$
  or a {\it star product}, if  
  for all $f,g \in \cC^\infty (X)$ the following holds:
  \begin{enumerate}[(DQ1)]
  \item 
    $f\star g = \sum\limits_{k\in \N}\, \mu_k (f,g) \, \lambda^k$,
    where the
    $\mu_k :\cC^\infty (X)\times \cC^\infty (X)\rightarrow \cC^\infty (X)$
    are bilinear maps and $\mu_0=\mu$ is the pointwise product on 
    $\cC^\infty (X)$,
  \item
    $[f,g]_\star -i \lambda \{ f,g\} \in\lambda^2 \cC^\infty (X)[[\lambda]]$,
    where 
    $[f,g]_\star $ is the commutator $f\star g - g \star f$,
  \item
    $f \star 1 = 1 \star f =f$.
  \end{enumerate}
  The deformation quantization is called {\it local}, if for all $k\in \N$
  \begin{equation}
    \supp \mu_k (f,g) \subset \supp f \cap \supp g,
  \end{equation}
  and {\it differential}, if all
  the $\mu_k$ are bidifferential operators on $X$. By a {\it bidifferential
  operator} on $X$ we hereby understand an operator 
  $\cC^\infty(X)\otimes \cC^\infty (X) \rightarrow \cC^\infty (X)$ which 
  in every 
  orbispace chart $(\tilde U,G,\varrho)$ is induced by a $G$-invariant 
  bidifferential on $\tilde U$.
\end{topic} 
\begin{example}
\label{ExQuantSympCone}
  Consider the symplectic cone $C = \R^2/ \Z_2$ of Example \ref{ExSympCone}.
  Let $\star$ be the Moyal--Weyl product on $\R^2$ that means 
  \begin{equation}
    f \star g = \sum_{k\in \N} \, \left( \frac{-i \lambda}{2} \right)^k
    \, \mu \big( \hat\Pi (f\otimes g)\big) \quad 
    \text{for all $f,g \in \cC^\infty (\R^2)$},
  \end{equation}
  where $\hat\Pi (f \otimes g) = 
  \frac{\partial}{\partial q} f \otimes \frac{\partial}{\partial p} g - 
  \frac{\partial}{\partial p} f \otimes \frac{\partial}{\partial q} g$
  and $\mu (f\otimes g) = f \, g$.
  Since the operator $\hat\Pi$ is $\Z_2$-invariant, 
  $\star$ can be restricted to an associative product on the space 
  $\cC^\infty (\R^2)^{\Z_2} [[\lambda]]$, where 
  $\cC^\infty (\R^2)^{\Z_2}$ denotes the algebra of $\Z_2$-invariant smooth
  functions on $\R^2$. But $\cC^\infty (\R^2)^{\Z_2}$ is canonically
  isomorphic to $\cC^\infty (C)$, hence we obtain a star product for $C$.
\end{example}


%
%
\section{Fedosov's quantization for orbispaces}
\label{SecFedQOrbi}
\begin{topic}
  In this section we will show how Fedosov's method for the construction
  of a (differentiable) star-product can be transferred to the arena of 
  orbispaces.
  The essential point hereby is to check that all of Fedosov's constructions 
  can be performed in a manner which is 
  natural with respect to morphisms of orbispace charts and invariant
  with respect to the involved symmetries. 
  We proceed analogously to {\sc Fedosov} \cite[Chap.~5]{Fed:DQIT} 
  (cf.~also \cite[Sec.~21]{CanWei:GMNA}).
  In particular, we will define the Weyl algebra bundle $\mathbb W X$
  of a symplectic orbispace $X$ and then construct a flat connection $D$
  on the Weyl algebra bundle such that the space of formal power series  
  in $\cC^\infty (X)$ can be (linearly) identified with the subalgebra
  of flat sections of $\mathbb W X$. Via this identification,
  $\cC^\infty (X)$ then inherits a star-product from $\mathbb W X$.
\end{topic}
\begin{topic}
 Let $V$ be a finite dimensional Poisson vector space and let $\Pi$
 ist Poisson bivector. One can then associate to $V$ the 
 {\it formal Weyl algebra} $\mathbb W V$ and the 
 {\it completed formal Weyl algebra} $\widehat{\mathbb W} V$ as follows. 
 As a (complex) vector space $\mathbb W V$ coincides 
 with $\operatorname{Sym}^\bullet (V^*) [[\lambda]]$, 
 the space of formal power series in $\lambda$ with coefficients in the 
 algebra of complex valued polynomial functions on $V$.  
 The coompleted formal Weyl algebra $\widehat{\mathbb W} V$ has
 $\widehat{\operatorname{Sym}}^\bullet (V^*) [[\lambda]]$
 as underlying linear space.
 Note that 
 $\operatorname{Sym}^\bullet (V^*)=
  \bigoplus_{s\in \N}\operatorname{Sym}^s (V^*)$
 is a graded algebra, where the product is given by $\mu$, the pointwise 
 product of functions, and the homogeneous component 
 $\operatorname{Sym}^s (V^*)$ consists of $s\,$-homogeneous
 polynomials. The profinite dimensional vector space
 $\widehat{\operatorname{Sym}}^\bullet (V^*)$
 coincides with $\prod_{s\in \N}\operatorname{Sym}^s (V^*)$ 
 and carries a natural descending filtration given by the powers
 $\widehat{\mathfrak m}^n$, where $ \widehat{\mathfrak m}$ is the kernel 
 of the canonical morphism 
 $\widehat{\operatorname{Sym}}^\bullet (V^*) \rightarrow \C \cong
 \operatorname{Sym}^0 (V^*)$. Moreover,
 $\widehat{\operatorname{Sym}}^\bullet (V^*)$ is complete 
 with respect to the topology defined by this filtration.

 By construction, $\mathbb W V$ is a subspace of $\widehat{\mathbb W} V$. 
 Every element $a \in \widehat{\mathbb W} V$ now has a unique
 representation of the form
 \begin{equation}
 \label{Eq4.1}
   a = \sum_{k\in \N, s\in\N}\, a_{sk} \lambda^k,
 \end{equation}
 where $a_{sk} \in \operatorname{Sym}^s (V^*)$ and where only finitely many
 $a_{sk}$ do not vanish for fixed $k$, if $a \in \mathbb W V$.
 Next recall that the Poisson bivector $\Pi$ can be written as 
 a finite sum $\Pi =\sum_i \,\Pi_{1i} \otimes \Pi_{2i}$
 with $\Pi_{1i},\Pi_{2i} \in V$
 and that the elements of $V$ act by derivations on 
 $\operatorname{Sym}^\bullet (V^*)$. 
 Therefore, the operator
 \begin{displaymath}
   \hat{\Pi} : \operatorname{Sym}^\bullet (V^*)\otimes_{\C}
   \operatorname{Sym}^\bullet (V^*) \rightarrow 
   \operatorname{Sym}^\bullet (V^*)\otimes_{\C}
   \operatorname{Sym}^\bullet (V^*) ,\quad f\otimes g \mapsto 
   \sum_i\, \Pi_{1i} f\otimes \Pi_{2i} g
 \end{displaymath}
 is well-defined and continuous with respect to the Krull topology
 defined by $\mathfrak m$.
 Hence, $\hat{\Pi}$ can be extended by $\C[[\lambda]]$-linearity and
 continuity 
 to an operator on $\widehat{\mathbb W} V \otimes_\C \widehat{\mathbb W} V$.
 The {\it Moyal--Weyl} product on $\widehat{\mathbb W} V$ then is given 
 as follows:
 \begin{equation}
   a \circ b := \mu \big( \exp (-i\lambda \hat{\Pi}) (a\otimes b) \big) :=
   \sum_{k\in\N} \, \frac{(-i\lambda)^k}{k!}\, \mu 
   \big( \hat{\Pi}^k (a\otimes b ) \big)
   \quad \text{for $a,b\in \widehat{\mathbb W}V$}.
 \end{equation}
 Thus $(\widehat{\mathbb W} V,\circ)$ becomes an associative algebra,
 and $\mathbb W V$ a subalgebra. The (completed) formal Weyl algebra 
 carries a (descending) filtration $(\widehat{\mathbb W}_n V)_{n\in \N}$ 
 defined by the {\it Fedosov-degree}
 \begin{equation}
   \deg_{\mathrm F} (a) = \min \{ s+2k \mid a_{sk} \neq 0\}, 
   \quad a\in \mathbb WV.
 \end{equation}
 This means that the subalgebra $\widehat{\mathbb W}_n V$ is given 
 by $\{ a\in \widehat{\mathbb W} V \mid \deg_{\mathrm F}(a) \geq n\}$.
 
 Additionally to $\widehat{\mathbb W} V$ we consider the algebra 
 $\Lambda^\bullet \widehat{\mathbb W} V := \Lambda^\bullet V 
 \otimes_\R \widehat{\mathbb W}V$
 of alternating forms with values in $\widehat{\mathbb W}V$. 
 The product $\circ$ on
 $\widehat{\mathbb W}V$ and the exterior product on 
 $\Lambda^\bullet V$ induce a
 product on $\Lambda^\bullet \widehat{\mathbb W}V$, denoted by $\circ$ 
 as well.
 Moreover, the filtration of $\widehat{\mathbb W}V$ induces a filtration 
 of $\Lambda^\bullet \widehat{\mathbb W} V$.

\end{topic}

 The following  result is crucial for all our further considerations.
 As the proof is obvious, we leave it to the reader. 
 \begin{proposition}
 \label{Prop4.3}
   Associate to every finite dimensional Poisson vector space $V$ 
   the completed formal Weyl algebra
   $\widehat{\mathbb W} V$ and to every linear Poisson map
   $f : W \rightarrow V$ the linear map 
   \begin{equation}
     \widehat{\mathbb W} f : \widehat{\mathbb W} V \rightarrow 
     \widehat{\mathbb W} W ,\quad
     a = \sum_{k\in \N, s\in\N}\,  a_{sk} \lambda^k\mapsto 
     \sum_{k\in \N, s\in\N}\,  f^* (a_{sk}) \lambda^k. 
   \end{equation}
   Then, $\widehat{\mathbb W}$ is a contravariant functor 
   with values in the category of profinite dimensional vector spaces.
   Likewise, $\Lambda^\bullet \widehat{\mathbb W}$ can be regarded 
   as a functor defined on the category of finite dimensional Poisson
   vector spaces with 
   values in the category of profinite dimensional vector spaces.
 \end{proposition}
 
\begin{topic} 
 Next let us consider a symplectic orbispace 
 $\big(X, (\omega_{\tilde U})\big)$. 
 Without loss of generality we can assume
 that every $\tilde U$ appearing as an index of 
 $(\omega_{\tilde U})$ is an orbispace chart 
 of some orbispace functor $\mathsf X$ such that
 $(\omega_{\tilde U})$ is an open $G$-invariant subset of $\R^{2n}$, 
 such that $G$ acts by linear symplectic maps on $\R^{2n}$  and finally 
 such that the symplectic form
 $\omega_{\tilde U}$ is given by 
 $\sum_{j=1}^n d\tilde x_j\wedge d\tilde x_{n+j}$,
 where $(\tilde x_1,\cdots ,\tilde x_{2n})$ are the natural coordinate
 functions over $\tilde U\subset \R^{2n}$.
 Given an element $(\tilde U, G, \varrho) \in \cU$, every fiber of
 $T\tilde U$ is a Poisson vector space, so we can apply 
 $\widehat{\mathbb W}$ fiberwise 
 and thus obtain the Weyl algebra bundle $\widehat{\mathbb W} \tilde U$.
 Likewise, the bundle of forms of the Weyl algebra 
 $\Lambda^\bullet \widehat{\mathbb W} \tilde U$ is constructed.
 Following {\sc Fedosov} \cite[Chap.~5]{Fed:DQIT} we will now introduce a 
 convenient representation of the sections  
 of these bundles. Let $(d\tilde x_1,\cdots ,d \tilde x_{2n})$ 
 be the local frame of  $T^*\tilde U$ corresponding to the 
 coordinates $(\tilde x_1,\cdots ,\tilde x_{2n})$ and denote by 
 $\tilde y_j$ for $j=1,\cdots,2n$ the canonical image of $d\tilde x_j$ 
 in the sectional space 
 $\Gamma^\infty (\operatorname{Sym}^\bullet(T^*\tilde U))$. Hereby, 
 $\operatorname{Sym}^\bullet$ is regarded as a fiberwise acting 
 functor on the category of finite dimensional vector bundles. As a 
 (toplogical) $\cC^\infty (\tilde U)$-module, 
 $\Gamma^\infty (\operatorname{Sym}^\bullet(T^*\tilde U))$ is 
 generated by the sections 
 $\tilde y^\alpha =\tilde y_1^{\alpha_1}\cdot\ldots\cdot\tilde
 y_n^{\alpha_n}$, where $\alpha \in \N^n$. 
 With these  notational agreements, a section 
 $a_{\tilde U} \in  \Gamma^\infty 
 (\Lambda^\bullet\widehat{\mathbb W} \tilde U)$ 
 resp.~an element 
 $a_{\tilde x} \in \Gamma^\infty (\widehat{\mathbb W} \tilde U)$
 (with $\tilde x$ denoting the footpoint)
 can be represented in the form
 \begin{equation}
   a_\Diamond = \sum_{k\in\N, \alpha \in \N^{2n}, l \in \N}\,
   \sum_{1\leq j_1< \cdots < j_l\leq 2n} \, 
   a_{\Diamond, k \alpha j_1\cdots j_l} \, 
   \tilde y^\alpha \, d\tilde x_{j_1} \wedge \cdots \wedge
   d\tilde x_{j_l} \, \lambda^k
 \end{equation}
 where $\Diamond$ is one of the symbols  $\tilde U $ or $\tilde x$, and
 the elements $a_{\tilde U,k \alpha j_1\cdots j_l}\in\cC^\infty (\tilde U)$  
 resp.~$a_{\tilde x,k \alpha j_1\cdots j_l}\in \C$ are uniquely defined.
 To simplify notation we  write $a_{\tilde x}$ not only
 for an element of $\widehat{\mathbb W} \tilde U$ with footpoint 
 $\tilde x$  but also for the evaluation of a section
 $a_{\tilde U}\in\Gamma^\infty (\Lambda^\bullet\widehat{\mathbb W} \tilde U)$ 
 at  $\tilde x$. 
\end{topic}
\begin{topic}
 In the following step we will lift the $G$-action to 
 $\widehat{\mathbb W} \tilde U$.
 Denote by $l_g $ the action of some group element $g$ on $\tilde U$.
 Then the derivative $T_{\tilde x}l_g$ is 
 a linear Poisson map, so by Proposition \ref{Prop4.3}
 \begin{displaymath}
   G\times \widehat{\mathbb W} \tilde U \rightarrow 
   \widehat{\mathbb W}\tilde U, \quad
   (g,a_{\tilde x})\mapsto \widehat{\mathbb W} 
   (T_{g \tilde x}l_{g^{-1}})(a_{\tilde x})
 \end{displaymath}
 is a $G$-action on $\widehat{\mathbb W}\tilde U$.
 Given a second element  $(\tilde V, H,\upsilon) \in \cU$ and 
 a morphism of orbispace charts 
 $(\varphi,\iota):(\tilde V,H,\upsilon)\rightarrow (\tilde U, G,\varrho)$ 
 the pair
 \begin{displaymath}
   (\widehat{\mathbb W} \varphi,\iota) : 
   (\widehat{\mathbb W} \tilde V,H)\rightarrow 
   (\widehat{\mathbb W} \tilde U, G), \quad (a_{\tilde y}, h) \mapsto 
   (\widehat{\mathbb W} (T_{\tilde y} \varphi)^{-1} \, (a_{\tilde y}),\iota) 
 \end{displaymath}
 induces a morphism in the category of profinite dimensional 
 vector spaces with symmetries, since by Proposition \ref{Prop4.3}
 \begin{equation}
   \begin{split}
     \widehat{\mathbb W} \varphi ( h a_{\tilde y} ) \, &  = 
     \widehat{\mathbb W} (T_{h \tilde y} \varphi)^{-1} \, (h a_{\tilde y}) = 
     \widehat{\mathbb W} (T_{h \tilde y}l_{h^{-1}}\circ 
     (T_{h \tilde y} \varphi)^{-1} )\,
     ( a_{\tilde y})  = \widehat{\mathbb W} (T_{\tilde y} 
     (\varphi \circ l_h))^{-1} 
     ( a_{\tilde y}) \\ & = 
     \widehat{\mathbb W} (T_{\tilde y}(l_{\iota (h)}\circ\varphi))^{-1} 
     ( a_{\tilde y}) = \widehat{\mathbb W} (T_{\iota (h) \, \varphi (\tilde y)}
     l_{\iota(h)^{-1}} ) \, \widehat{\mathbb W} \varphi ( a_{\tilde y})
     = \iota (h)\, \widehat{\mathbb W} \varphi  (a_{\tilde y}).
   \end{split}
 \end{equation}
 Thus we obtain an orbibundle functor $\widehat{\mathbb W} \mathsf X$
 which associates to every $\tilde U$ the pair 
 $(\widehat{\mathbb W} \tilde U, G)$ 
 and to every morphism $(\varphi,\iota)$ between elements of $\cU$ 
 the morphism $(\widehat{\mathbb W} \varphi,\iota)$. 
 The functor $\widehat{\mathbb W} \mathsf X$ induces a 
 vector orbibundle $\widehat{\mathbb W} X \rightarrow X $, called the 
 {\it Weyl algebra orbibundle} of $X$, and 
 an orbibundle atlas 
 $(\widehat{\mathbb W} \tilde U,G,\widehat{\mathbb W} \varrho)$. 
 Likewise, one constructs the vector 
 orbibundle $\Lambda^\bullet \widehat{\mathbb W} X \rightarrow X$
 of so-called {\it forms of the Weyl algebra orbibundle}.
 By construction, the orbibundles $\widehat{\mathbb W} X $ and 
 $\Lambda^\bullet \widehat{\mathbb W} X $ are reduced, hence 
 Remark \ref{RemSecRedOrbiBdl} applies to sections of
 $\widehat{\mathbb W} X $ and $\Lambda^\bullet \widehat{\mathbb W} X$.
 \end{topic}
 \begin{proposition}
   The sectional spaces 
   $\Gamma_{\text{\sf \tiny str}}^\infty (\widehat{\mathbb W} X)$ and 
   $\Gamma_{\text{\sf \tiny str}}^\infty 
   (\Lambda^\bullet \widehat{\mathbb W} X)$ carry in a natural 
   way a $\C[[\lambda]]$-bilinear associative product $\circ$
   such that 
   \begin{equation}
   \label{Eq4.8}
     (a\circ b)_{\tilde U} = a_{\tilde U} \circ b_{\tilde U}
     \quad \text{for all 
     $a,b \in \Gamma_{\text{\sf \tiny str}}^\infty (\widehat{\mathbb W} X)$
     (resp.~$a,b\in \Lambda^\bullet 
     \Gamma_{\text{\sf \tiny str}}^\infty (\widehat{\mathbb W} X)$)}.
   \end{equation}
   Moreover, the space 
   $\Gamma_{\text{\sf \tiny str}}^\infty 
   (\Lambda^\bullet \widehat{\mathbb W} X)$ thus 
   becomes a graded and filtered algebra, where the graduation degree is 
   given by the form degree and the filtration degree by the Fedosov degree.
   The topology defined by the Fedosov filtration  provides
   $\Gamma_{\text{\sf \tiny str}}^\infty 
   (\Lambda^\bullet \widehat{\mathbb W} X)$ 
   with the structure of a complete topological vector space.
 \end{proposition}
   The graded commutator on 
   $\Gamma_{\text{\sf \tiny str}}^\infty
   (\Lambda^\bullet \widehat{\mathbb W} X)$
   with respect to the product $\circ$ will be denoted by $[\cdot,\cdot]$.
   \vspace{2mm} \\
 \begin{proof}
   Using the $G$-invariance of the symplectic form $\omega_{\tilde U}$ it is 
   straightforward to check that 
   \begin{equation}
   \label{Eq4.9}
     g a_{\tilde U} \circ g b_{\tilde U} = g ( a_{\tilde U} \circ b_{\tilde U})
     \quad \text{for all 
     $a_{\tilde U},b_{\tilde U} \in 
     \Gamma^\infty (\widehat{\mathbb W}\tilde U)$}.
   \end{equation}
   Moreover, if $(\varphi,\iota) $ is a morphism like above, then
   \begin{equation}
     \widehat{\mathbb W} \varphi ( a_{\tilde y} \circ b_{\tilde y} ) =
     \widehat{\mathbb W} \varphi ( a_{\tilde y}) \circ
     \widehat{\mathbb W} \varphi ( b_{\tilde y} ) \quad
     \text{for all $a_{\tilde y},b_{\tilde y}
     \in \widehat{\mathbb W} \tilde V $ and $\tilde y \in\tilde V$}.
   \end{equation}
   Hence, Eq.~(\ref{Eq4.8}) defines a section 
   $a\circ b\in\Gamma_{\text{\sf \tiny str}}^\infty (\widehat{\mathbb W} X)$. 
   From the corresponding 
   properties of the product on 
   $\Gamma^\infty (\widehat{\mathbb W}\tilde U)$ one now 
   concludes that $\circ$ is a $\C[[\lambda]]$-bilinear associative product.
   The same argument proves that $\circ$ is a product on 
   $\Gamma_{\text{\sf \tiny str}}^\infty 
   (\Lambda^\bullet \widehat{\mathbb W}X)$.
   The remaining part of the claim is obvious.
 \end{proof}

 \begin{topic}
 Let us now choose a symplectic connection 
 $(\nabla_{\tilde U})$ on $X$
 and extend it in a natural way to a connection on 
 $\Lambda^\bullet \widehat{\mathbb W} X$ by putting 
 \begin{equation}
   (\nabla b)_{\tilde U} = \sum_{j=1}^{2n}\, \sum_{k,\alpha,l}\,
   \sum_{1\leq j_1 <\cdots <j_l\leq 2n}\,
   \nabla_{\tilde U,\frac{\partial}{\partial \tilde x_j}}  
   \big( b_{\tilde U, k \alpha j_1\cdots j_l} \, 
   \tilde y^\alpha \big) \, d\tilde x_j \wedge d\tilde x_{j_1} \wedge 
   \cdots \wedge  d\tilde x_{j_l} \, \lambda^k.
 \end{equation} 
 By construction, $(\nabla b)_{\tilde U}$ is a $G$-equivariant 
 section of $\Lambda^\bullet \widehat{\mathbb W} \tilde U$, and 
 $\varphi^* (\nabla b)_{\tilde U} = (\nabla b)_{\tilde V}$ holds for every
 morphism 
 $(\varphi,\iota) : (\tilde V,H,\upsilon) \rightarrow (\tilde U,G,\varrho)$.
 Hence, the family $\big((\nabla b)_{\tilde U}\big)$ gives rise to a 
 section of
 $\Lambda^\bullet \widehat{\mathbb W} X$, and the connection
 $\nabla : \Gamma_{\text{\sf \tiny str}}^\infty 
 (\Lambda^\bullet \widehat{\mathbb W} X) 
 \rightarrow \Gamma_{\text{\sf \tiny str}}^\infty 
 (\Lambda^\bullet \widehat{\mathbb W} X) $ is 
 well-defined.  Over $\tilde U$, the components of
 $\nabla b$ are given by
 \begin{equation}
   (\nabla b)_{\tilde U} = d b_{\tilde U} + \frac{i}{\lambda} 
   [\Gamma_{\tilde U},b_{\tilde U}],
 \end{equation}
 where $\Gamma_{\tilde U} = \frac 12 \sum_{i,j,k} \Gamma_{\tilde U, ijk}
 \tilde y_i \tilde y_j \, d\tilde x_k$ is a local one-form 
 and the $\Gamma_{\tilde U, ijk}$ are the Christoffel symbols
 of $\nabla$, i.e.~$\nabla_{\tilde U, \frac{\partial}{\partial \tilde x_i}}
 \frac{\partial}{\partial \tilde x_j} = \sum_{k,l} \Gamma_{\tilde U, ijk} 
 \omega_{kl} \frac{\partial}{\partial \tilde x_l}$. Moreover, 
 the family $R=(R_{\tilde U})$ with 
 $R_{\tilde U} = d\Gamma_{\tilde U} + 
 \frac 12 [\Gamma_{\tilde U},\Gamma_{\tilde U}]$
 defines a smooth section of $\Lambda^2\widehat{\mathbb W} X$. 
 From \cite[Lem.~5.1.3]{Fed:DQIT} one concludes that 
 \begin{equation}
   \nabla^2 b = \frac{i}{\lambda} [R,b] \quad 
   \text{for all 
   $b\in\Gamma_{\text{\sf \tiny str}}^\infty 
   (\Lambda^\bullet\widehat{\mathbb W}X)$}.
 \end{equation}
 Hence, $R$ can be interpreted as the {\it curvature form} of $\nabla$.

 We will now employ Fedosov's idea and construct a flat
 connection $D$ on $\Lambda^\bullet \widehat{\mathbb W}X$ of the form
 \begin{equation}
   Db = \nabla b + \delta b + \frac{i}{\lambda} [ r,b]
   \quad 
   \text{for all $b\in \Gamma_{\text{\sf \tiny str}}^\infty 
   (\Lambda^\bullet \widehat{\mathbb W}X)$},
 \end{equation}
 where
 $r\in \Gamma_{\text{\sf \tiny str}}^\infty 
 (\Lambda^1 \widehat{\mathbb W}X)$ and  
 $\delta:\Gamma_{\text{\sf \tiny str}}^\infty 
 (\Lambda^\bullet \widehat{\mathbb W}X) \rightarrow 
 \Gamma_{\text{\sf \tiny str}}^\infty 
 (\Lambda^\bullet \widehat{\mathbb W}X)$ is a graded 
 derivation which locally is defined by 
 \begin{equation}
 \label{Eq4.15}
   (\delta b)_{\tilde U} = \sum_k \, d\tilde x_k \wedge
   \frac{\partial b_{\tilde U}}{\partial \tilde y_k} 
   = - \frac{i}{\lambda} \sum_{k,l} \, [\omega_{kl} \, 
   \tilde y_k\, d\tilde x_l, b_{\tilde U}] .
 \end{equation}
 Note that Eq.~(\ref{Eq4.15}) gives rise to an operator on the space of 
 smooth stratified sections of 
 $\Lambda^\bullet \widehat{\mathbb W} X$ indeed, since 
 the $(\delta b)_{\tilde U}$ are $G$-equivariant and transform 
 naturally under morphisms of orbispace charts.
 Similarly one concludes that the operator
 $\delta^* :\Gamma_{\text{\sf \tiny str}}^\infty 
 (\Lambda^\bullet \widehat{\mathbb W}X) \rightarrow 
 \Gamma_{\text{\sf \tiny str}}^\infty 
 (\Lambda^\bullet \widehat{\mathbb W}X)$ is well-defined
 by putting locally
 \begin{equation}
   (\delta^*b)_{\tilde U} =\sum_k \, \tilde y_k \cdot \big(
   \frac{\partial}{\partial \tilde x_k} \, \llcorner \, b_{\tilde U}\big).
 \end{equation}
 Finally, $\delta^*$ gives rise to a third operator  
 $\delta^-:\Gamma_{\text{\sf \tiny str}}^\infty 
 (\Lambda^\bullet \widehat{\mathbb W}X) \rightarrow 
 \Gamma_{\text{\sf \tiny str}}^\infty 
 (\Lambda^\bullet \widehat{\mathbb W}X)$ by the local definition
 \begin{equation}
   (\delta^- b)_{\tilde U} = \sum_{q+l>0}\,
   \frac{1}{q+l} \, \delta^* (b_{\tilde U, ql}),
 \end{equation} where
   $b_{\tilde U, ql} = \sum\limits_{k, \,|\alpha|= q}\,
   \sum\limits_{1\leq j_1< \cdots < j_l\leq 2n} \, 
   b_{\tilde U, k \alpha j_1\cdots j_l} \, 
   \tilde y^\alpha \, d\tilde x_{j_1} \wedge \cdots \wedge
   d\tilde x_{j_l} \, \lambda^k$.
 \end{topic}
 The following propositions can now be easily deduced from the corresponding 
 ones in the smooth case.
 \begin{proposition}
   For every $b\in \Gamma_{\text{\sf \tiny str}}^\infty 
   (\Lambda^\bullet \widehat{\mathbb W}X )$
   one has the so-called {\it Hodge--de Rham decomposition} 
   \begin{equation}
     b = \delta \, \delta^- b + \delta^- \, \delta b + \sigma (b),
   \end{equation}
   where 
   $\sigma : \Gamma_{\text{\sf \tiny str}}^\infty 
   (\Lambda^\bullet \widehat{\mathbb W}X)\rightarrow 
   \cC^\infty (X)$, $(b_{\tilde U}) \mapsto (b_{\tilde U,00})$ 
   is the symbol map.
 \end{proposition}
 \begin{proof}
   This follows immediately from \cite[Lem.~5.1.2]{Fed:DQIT}.
 \end{proof}
 \begin{proposition}
   Given $r\in \Gamma_{\text{\sf \tiny str}}^\infty 
   (\Lambda^1 \widehat{\mathbb W}X)$ 
   let $\Omega$ be the two-form
   $-\omega + R-\delta r +\nabla r + \frac i\lambda r^2$ with $R$ the 
   curvature
   form of $\nabla$. Then $\Omega$ is the curvature form of    
   the connection
   $D = \nabla + \delta b + \frac{i}{\lambda} [r,\cdot]$ that means 
   $\Omega$ satisfies
   \begin{equation}
     D^2 b=  \frac i\lambda [\Omega,b] \quad 
     \text{for all $b\in \Gamma_{\text{\sf \tiny str}}^\infty (\Lambda^\bullet 
     \widehat{\mathbb W}X)$}.
   \end{equation}
 \end{proposition}
 \begin{proof}
   By \cite[Lem.~5.1.5]{Fed:DQIT}, the equality 
   $D^2_{\tilde U} b_{\tilde U}=\frac i\lambda 
   [\Omega_{\tilde U},b_{\tilde U}]$ holds true for all 
   $b_{\tilde U}\in \Gamma^\infty 
   (\Lambda^\bullet \widehat{\mathbb W}\tilde U)$, hence
   the claim follows.
 \end{proof}
 \begin{proposition}
   Given $r_0\in \Gamma_{\text{\sf \tiny str}}^\infty 
   (\Lambda^1 \widehat{\mathbb W}X)$ with 
   $\deg_{\mathrm F} (r_0)\geq 2$ there exists a unique 
   $r\in \Gamma_{\text{\sf \tiny str}}^\infty 
   (\Lambda^1 \widehat{\mathbb W}X)$ with 
   $\deg_{\mathrm F} (r)\geq \deg_{\mathrm F} (r_0)$ such that
   \begin{equation}
   \label{Eq4.19}
     r = r_0 + \delta^{-} \left( \nabla r + \frac{i}{\lambda} \, r^2 \right).
   \end{equation}
 \end{proposition}
 \begin{proof}
   Consider the operator 
   \begin{displaymath}
     K: \Gamma_{\text{\sf \tiny str}}^\infty 
     (\Lambda^1 \widehat{\mathbb W}_2X) \rightarrow
     \Gamma_{\text{\sf \tiny str}}^\infty 
     (\Lambda^1 \widehat{\mathbb W}X), \quad s \mapsto 
     r_0 + \delta^{-} \left( \nabla s + \frac{i}{\lambda} \, s^2 \right).
   \end{displaymath}
   It is immediate to check that $K$ has image in 
   $\Gamma_{\text{\sf \tiny str}}^\infty (\Lambda^1 \widehat{\mathbb W}_2X)$
   and that $K$ is contractible with respect to the Fedosov 
   filtration in the sense that
   \begin{displaymath}
     \deg_{\mathrm F} \big( K(s) - K(s') \big) >
     \deg_{\mathrm F} (s- s') \quad \text{for all 
     $s,s' \in \Gamma_{\text{\sf \tiny str}}^\infty 
     (\Lambda^1 \widehat{\mathbb W}_2X)$}. 
   \end{displaymath}
   Hence, since $\Gamma_{\text{\sf \tiny str}}^\infty 
   (\Lambda^1 \widehat{\mathbb W}_2X)$ is complete with respect to
   the topology given by the Fedosov filtration, one concludes 
   by a Banach fixed point type argument that
   there exists a unique $r$ satisfying the claim.
 \end{proof}
 \begin{corollary}
   Let $R$ be the curvature form of a symplectic connection $\nabla$ on $X$
   and $r_0= \delta^- R$. Then, if $r$ is the solution of
   (\ref{Eq4.19}), the curvature $\Omega$ of 
   $D = \nabla + \delta b + \frac{i}{\lambda} [r,\cdot]$ is a central element
   with respect to $\circ$ and satisfies $\Omega = - \omega$. 
   In particular, $D$ then is a flat connection.
 \end{corollary}
 \begin{proof}
   We follow the argument of \cite[Thm.~5.2.2]{Fed:DQIT}.
   First, note that $(\delta^-)^2=0$, so one has by the 
   Hodge--de Rham decomposition and Eq.~(\ref{Eq4.19}) 
   \begin{displaymath}
     \delta^- (\Omega + \omega ) = \delta^- \big( R - \delta r + \nabla r
     + \frac{i}{\lambda} r^2 \big) = r - \delta^- \delta r =
     \delta (\delta^-)^2 R = 0.
   \end{displaymath}
   Using again the Hodge--de Rham decomposition, the Bianchi identity
   $D \Omega =0$ and the equality $D\omega = d\omega =0$ entail that
   \begin{displaymath}
     \Omega + \omega = \delta^- (D+\delta) (\Omega +\omega).
   \end{displaymath}
   Now the operator 
   $\delta^- (D+\delta) = \delta^- (\nabla +\frac{i}{\lambda} \, [r,\cdot])$
   raises the Fedosov degree by $1$, hence one concludes that 
   $\Omega + \omega =0$. But this implies also that $\Omega$ is central,
   so the claim follows.
 \end{proof}
   For the flat connection $D$ constructed in the corollary let 
   $\widehat{\mathbb W}_D X$ 
   be the space of all flat sections, that means the space of all elements 
   $a \in \Gamma_{\text{\sf \tiny str}}^\infty (\widehat{\mathbb W}X)$ 
   satisfying
   $Da =0$. Then $\widehat{\mathbb W}_D X$ forms a subalgebra of 
   $\Gamma_{\text{\sf \tiny str}}^\infty 
   (\widehat{\mathbb W}X)$,
   since $D$ is a graded derivation with respect to $\circ$.
   Using the above results one now proves the following result 
   literally like Thm.~5.2.4 of \cite{Fed:DQIT}.
 \begin{theorem}
   Let $X$ be a symplectic orbispace, $\nabla$ a symplectic
   connection on $X$ and $D$ the flat connection on 
   $\Lambda^\bullet \widehat{\mathbb W} X$ defined above.
   Then the symbol map induces a linear isomorphism
   $\sigma : \widehat{\mathbb W}_D X \rightarrow \cC^\infty (X) [[\lambda]]$.
 \end{theorem}
 \begin{proof}
  Choose $f\in \cC^\infty (X)[[\lambda]]$ and consider the equation
  \begin{equation}
    \label{EqSymInv}
      s = f + \delta^- (D+\delta) s, \quad s \in 
      \Gamma_{\text{\sf \tiny str}}^\infty 
      (\widehat{\mathbb W}X) .
  \end{equation}
  Since the operator $s\mapsto f + \delta^- (D+\delta) s$ is contractible in
  the above stated sense, this equation has a unique solution $s$.
  Let us show that $s \in \widehat{\mathbb W}_D X$ and $\sigma (s) =f$.
  First check by the Hodge--de Rham decomposition that
  \begin{displaymath}
    \delta^- D s = s - f -\delta ^- \delta s =  \delta  \delta ^-s =0.
  \end{displaymath}
  Using the Hodge--de Rham decomposition again, one gets $\sigma (s) =f$. 
  Applying the Hodge--de Rham decomposition a third time, but 
  now to the argument $Ds$, one concludes by $D^2 =0$ and $\delta^- D s$
  that
  \begin{displaymath}
    Ds = \delta^- (D +\delta) Ds.
  \end{displaymath}
  But this  equation has a unique solution, namely $Ds=0$, since the
  operator $\delta^- (D +\delta) $ is contractible.
  Hence $s \in \widehat{\mathbb W}_D X$ and $\sigma (s) =f$.
  Conversely, every $s\in \widehat{\mathbb W}_D X$ with $\sigma (s) =f$
  satisfies (Eq.~\ref{EqSymInv}) by the Hodge--de Rham decomposition.
  Thus, the theorem follows.
 \end{proof}

   Denote by 
   $Q: \cC^\infty (X) [[\lambda]]  \rightarrow \widehat{\mathbb W}_D X$  
   the inverse of the symbol map or in other words the quantization map. 
   The theorem now entails our main result.
 \begin{corollary}
   Let $\star : \cC^\infty (X) [[\lambda]]\times \cC^\infty (X)[[\lambda]]
   \rightarrow \cC^\infty (X) [[\lambda]]$ be the uniquely
   determined $\C[[\lambda]]$-bilinear map such that 
   \begin{displaymath}
      f\star g = \sigma ( Q (f) \circ Q(g)) \quad 
      \text{for all  $f,g \in \cC^\infty (X)$}.    
   \end{displaymath}
   Then $\star$ is a star product for $X$.
 \end{corollary}
 \begin{corollary}
   Every symplectic orbifold possesses a star product.
 \end{corollary}


%
%
%
\newpage
\addcontentsline{toc}{section}{References}
\ifx\undefined\allcaps\def\allcaps#1{#1}\fi
\providecommand{\bysame}{\leavevmode\hbox to3em{\hrulefill}\thinspace}
\providecommand{\MR}{\relax\ifhmode\unskip\space\fi MR }
\providecommand{\MRhref}[2]{%
  \href{http://www.ams.org/mathscinet-getitem?mr=#1}{#2}
}
\providecommand{\href}[2]{#2}

\end{document}